\documentclass[prd,aps,floatfix,twocolumn]{revtex4-2}
\usepackage{amsfonts,amsmath,amssymb,amsthm}
\usepackage{bm}
\usepackage{booktabs}
\usepackage{braket}
\usepackage{cases}
\usepackage{color}
\usepackage{dcolumn}
\usepackage{epsfig}
\usepackage{epstopdf}   
\usepackage{graphicx}
\usepackage[colorlinks,citecolor=blue,anchorcolor=red,menucolor=red,linkcolor=red,filecolor=red,runcolor=red,urlcolor=blue,frenchlinks=red]{hyperref}
\usepackage{indentfirst}
\usepackage{latexsym}
\usepackage{longtable}
\usepackage{multirow}
\usepackage{rotating}
\usepackage{slashed}  
\usepackage{subfigure}

\allowdisplaybreaks[4]
\begin{document}
\title{Systematics of fully heavy tetraquarks}
\author{Xin-Zhen Weng$^1$}
\email{xzhweng@pku.edu.cn}
\author{Xiao-Lin Chen$^1$}
\email{chenxl@pku.edu.cn}
\author{Wei-Zhen Deng$^1$}
\email{dwz@pku.edu.cn}
\author{Shi-Lin Zhu$^{1,2,3}$}
\email{zhusl@pku.edu.cn}
\affiliation{
$^1$School of Physics and State Key Laboratory of Nuclear Physics and Technology, Peking University, Beijing 100871, China \\
$^2$Center of High Energy Physics, Peking University, Beijing 100871, China \\
$^3$Collaborative Innovation Center of Quantum Matter, Beijing 100871, China
}
\begin{abstract}

In this work, we systematically study the mass spectrum of the fully heavy tetraquark in an extended chromomagnetic model, which includes both color and chromomagnetic interactions.
%
Numerical results indicate that the energy level is mainly determined by the color interaction, which favors the color-sextet $\ket{(QQ)^{6_{c}}(\bar{Q}\bar{Q})^{\bar{6}_{c}}}$ configuration over the color-triplet $\ket{(QQ)^{\bar{3}_{c}}(\bar{Q}\bar{Q})^{3_{c}}}$ one.
%
The chromomagnetic interaction mixes the two color configurations and gives small splitting.
%
The ground state is always dominated by the color-sextet configuration.
%
%
We find no stable state below the lowest heavy quarkonium pair thresholds.
Most states may be wide since they have at least one $S$-wave decay channel into two $S$-wave mesons.
One possible narrow state is the $1^{+}$ $bb\bar{b}\bar{c}$ state with a mass $15719.1~\text{MeV}$.
It is just above the $\eta_{b}\bar{B}_{c}$ threshold.
%
%
But this channel is forbidden because of the conservation of the angular momentum and parity.
%

\end{abstract}

\maketitle
\thispagestyle{empty} 

\section{Introduction}
\label{Sec:Introduction}

In the quark model~\cite{GellMann:1964nj,Zweig:1964pd}, a normal hadron is composed of quark-antiquark pair (meson) or three quarks (baryons).
%
The exotic states consisting of more than three quarks have not been observed until this century.
In 2003, the Belle Collaboration measured the exclusive $B^{\pm}{\to}K^{\pm}\pi^{+}\pi^{-}J/\psi$ decays, and observed the first charmoniumlike state $X(3872)$~\cite{Choi:2003ue}, whose quantum number was later determined to be $I^{G}J^{PC}=0^{+}1^{++}$~\cite{Zyla:2020zbs}.
%
Since then, lots of charmoniumlike and bottomoniumlike states are found, such as the $Y(4260)$~\cite{Aubert:2005rm}, $Z_{c}(3900)$~\cite{Ablikim:2013mio,Liu:2013dau}, $Z_{b}(10610)$ and $Z_{b}(10650)$~\cite{Bondar:2011aa} states.
Various theoretical interpretations, such as the molecule~\cite{Swanson:2003tb,Carames:2010zz,Chen:2015add}, the compact tetraquark~\cite{Cui:2006mp,Park:2013fda}, the hybrid meson~\cite{Zhu:2005hp,Esposito:2016itg}, etc. have been proposed to explain their nature.
More details can be found in Refs.~\cite{Chen:2016qju,Esposito:2016noz,Lebed:2016hpi,Ali:2017jda,Guo:2017jvc,Yuan:2018inv,Brambilla:2019esw,Liu:2019zoy} and references therein.

Another interesting structure is the fully heavy tetraquarks.
%
In 2016, the CMS collaboration reported the first observation of the $\Upsilon(1S)$ pair production in $pp$ collisions at $\sqrt{s}=8~\text{TeV}$~\cite{Khachatryan:2016ydm}.
They found an exotic structure in $\mu^+\mu^-\mu^+\mu^-$ channel with a global significance of $3.6\sigma$.
Its invariant mass is $18.4\pm0.1(\text{stat.})\pm0.2(\text{syst.})~\text{GeV}$~\cite{Durgut:2018:ukn001}.
Two years later, the LHCb collaboration searched for the $X_{b\bar{b}b\bar{b}}$ in the $\Upsilon(1S)\mu^+\mu^-$ final state, but they did not see any significant excess in the range $17.5\sim20.0~\text{GeV}$~\cite{Aaij:2018zrb}.
Last year, the {A$_N$DY} collaboration investigated the dijet mass in Cu+Au collisions, and found a peak at $M=18.12\pm0.15(\text{stat.})\pm0.6(\text{syst.})~\text{GeV}$, which is possibly an all-$b$ tetraquark~\cite{Bland:2019aha}.
In the full-charm sector, the charmonium pair production had also been observed by the NA3~\cite{Badier:1982ae,Badier:1985ri}, D0~\cite{Abazov:2014qba}, LHCb~\cite{Aaij:2014rms} and Belle~\cite{Yang:2014yyy} collaborations.
Very recently, the LHCb collaboration observed a narrow structure and a wide structure in the $J/\psi$-pair invariant mass spectrum in the range of $6.2\sim7.2~\text{GeV}$, which could be all-charm hadrons~\cite{Aaij:2020fnh}.
%

%
%
Theoretically, the fully heavy tetraquarks have been studied with various methods, such as 
the potential quark model~\cite{Bai:2016int,Karliner:2016zzc,Wu:2016vtq,Debastiani:2017msn,Richard:2017vry,Esposito:2018cwh,Bedolla:2019zwg,Liu:2019zuc,Wang:2019rdo,Braaten:2020nwp,Deng:2020iqw,Giron:2020wpx,Jin:2020jfc,Lu:2020cns},
the QCD sum rules~\cite{Chen:2016jxd,Wang:2017jtz,Chen:2018cqz,Wang:2018poa,Chen:2020xwe},
the covariant Bethe-Salpeter equations~\cite{Heupel:2012ua}
and the lattice QCD~\cite{Hughes:2017xie}.
In Ref.~\cite{Debastiani:2017msn}, Debastiani and Navarra found that the lowest $S$-wave  $cc\bar{c}\bar{c}$ tetraquarks might be below the di-charmonium thresholds.
Anwar {\it et~al.}~\cite{Anwar:2017toa} studied the $bb\bar{b}\bar{b}$ tetraquarks with a nonrelativistic effective field theory (NREFT) at the leading order (LO) and a relativized diquark model.
The ground state masses in the two approaches are $18.72\pm0.02~\text{GeV}$ and $18.75~\text{GeV}$ respectively, which indicates the existence of a $bb\bar{b}\bar{b}$ bound state below the $\eta_{b}\eta_{b}$ threshold.
However, many other studies suggest that the ground state of fully heavy tetraquarks is above the di-meson threshold.
For example, Wang {\it et~al.}~\cite{Wang:2019rdo} used two quark models with different potentials to study the $QQ\bar{Q}'\bar{Q}'$ tetraquarks.
They found that the $QQ\bar{Q}'\bar{Q}'$ tetraquarks are always above the $\eta_{Q}\eta_{Q}$/$B_{c}B_{c}$ thresholds and no bound tetraquark states exist.
In Ref.~\cite{Hughes:2017xie}, Hughes {\it et~al.} adopted the first-principles lattice nonrelativistic QCD methodology to study the lowest energy eigenstate of the $bb\bar{b}\bar{b}$ systems with quantum numbers $0^{++}$, $1^{+-}$ and $2^{++}$, and did not find any state below the lowest bottomonium-pair threshold.

In the quark model~\cite{Eichten:1978tg,Isgur:1977ef,Godfrey:1985xj,Capstick:1986bm,DeRujula:1975qlm},
the hadron mass can be decomposed into the quark masses, the kinetic energy and the potential interaction.
Usually, the potential includes the color-independent Coulomb and confinement interactions, and the hyperfine interactions like the spin-spin, spin-orbit, and tensor terms.

When restricted to the ground state, one can use the simplified chromomagnetic model~\cite{Sakharov:1966tua,DeRujula:1975qlm,Jaffe:1976ig,Jaffe:1976ih,Cui:2006mp,Buccella:2006fn,Wu:2016vtq}.
In this model, the quark masses, the kinetic energy and the spin-independent interaction are absorbed into the effective quark masses, and the spin-spin interaction is simplified to be the chromomagnetic interaction.
The spin-orbit and tensor interactions are ignored since we only consider the $S$-wave states.
This simplified model can well explain the hyperfine splittings of ordinary hadrons.
However, the recent studies indicate that the one-body effective quark masses are not enough to account for all of the two-body spin-independent color interaction.
For example, Karliner {\it et~al.}~\cite{Karliner:2014gca} found that a color-related interaction between heavy (anti-)quark and strange (or heavy) quark should be introduced to account for the heavy meson and baryon masses.
Then they used this model to predict the $\Xi_{cc}$ mass $M_{\Xi_{cc}}=3627\pm12~\text{MeV}$, which is very close to the LHCb's measurement~\cite{Aaij:2017ueg}.

In Ref.~\cite{Weng:2018mmf}, we generalized the chromomagnetic model by including the color interaction in the Hamiltonian.
For the color-singlet hadrons, we found that the effective quark masses can be absorbed into the color interaction.
With this model, we can reproduce the mass of doubly charm baryon $\Xi_{cc}$~\cite{Weng:2018mmf} and the recently observed $P_{c}$ pentaquarks~\cite{Weng:2019ynv}.
In this work, we will use the extended chromomagnetic model to study the spectrum of the $S$-wave fully heavy tetraquarks.
With the wave function obtained, we can also estimate their decay properties.

This work is organized as follows.
In Sec.~\ref{Sec:Model}, we introduce the extended chromomagnetic model, and present the wave function bases of the fully heavy tetraquarks.
Then we discuss the numerical results in Sec.~\ref{Sec:Result}.
In Sec.~\ref{Sec:Conclusion} we give a brief summary.
%

\section{The Extended Chromomagnetic Model}
\label{Sec:Model}


In the chromomagnetic model, the mass of the $S$-wave hadron consists of the effective quark masses and chromomagnetic (CM) interaction~\cite{Cui:2006mp,Buccella:2006fn,Wu:2016vtq,Liu:2019zoy}
\begin{equation}\label{eqn:cm}
H=
\sum_{i}m_{i}
-
\sum_{i<j}v_{ij}
\bm{S}_{i}\cdot\bm{S}_{j}
\bm{F}_{i}\cdot\bm{F}_{j}\,,
\end{equation}
where $\bm{S}_{i}=\bm{\sigma}_i/2$ ($\bm{F}_{i}={\bm{\lambda}}_i/2$) is the quark spin operator (color operator).
%
For the antiquark, $\bm{S}_{\bar{q}}=-\bm{S}_{q}^{*}$ and $\bm{F}_{\bar{q}}=-\bm{F}_{q}^{*}$.
The $m_i$ is the $i$th quark's (or antiquark's) effective mass.
And the $v_{ij}$ is effective coupling constant
\begin{equation}\label{eqn:CM:coupling-constant}
v_{ij}=
\frac{8\pi}{3m_i m_j}
\Braket{\alpha_s(r)\delta^3(\bm{r})}\,,
\end{equation}
which depends on the constituent quark masses and the spatial wave function.

Comparing to the dynamical quark model~\cite{Godfrey:1985xj,Capstick:1986bm}, we see that there is an additional color interaction, which contributes to the mass of the $S$-wave hadron, and is impossible to be absorbed into the effective quark masses or the chromomagnetic interaction.
This is also supported by some concrete studies concerning baryons~\cite{Karliner:2014gca} and tetraquarks~\cite{Hogaasen:2013nca}.
Here we introduce a colorelectric term into the model~\cite{Hogaasen:2013nca,Weng:2018mmf}
\begin{equation}\label{eqn:CE}
H_{\text{CE}}
=
-\sum_{i<j}A_{ij}\bm{\lambda}_i\cdot\bm{\lambda}_j\,.
\end{equation}
%
%
Since
\begin{eqnarray}\label{eqn:m+color=color}
&&\sum_{i<j}
\left(m_i+m_j\right)
\bm{F}_{i}\cdot\bm{F}_{j}
\notag\\
&&\quad=
\left(\sum_{i}m_{i}\bm{F}_i\right)\cdot\left(\sum_{i}\bm{F}_{i}\right)
-
\frac43\sum_{i}m_{i} \,,
\end{eqnarray}
and the total color operator $\sum_i\bm{F}_i$ nullifies any color-singlet physical state,
we can rewrite the model Hamiltonian as~\cite{Weng:2018mmf,Weng:2019ynv}
\begin{equation}\label{eqn:hamiltonian:final}
H=
-\frac{3}{4}
\sum_{i<j}m_{ij}V^{\text{C}}_{ij}
-
\sum_{i<j}v_{ij}V^{\text{CM}}_{ij} \,,
\end{equation}
where the quark pair mass parameter is
\begin{equation}\label{eqn:para:color+m}
m_{ij}
=
\left(m_i+m_j\right)+\frac{16}{3}A_{ij} \,,
\end{equation}
and the color and CM interactions between quarks are defined as
\begin{eqnarray}
V^{\text{C}}_{ij}
&=&
\bm{F}_{i}\cdot\bm{F}_{j} \,, \\
V^{\text{CM}}_{ij}
&=&
\bm{S}_{i}\cdot\bm{S}_{j}
\bm{F}_{i}\cdot\bm{F}_{j} \,.
\end{eqnarray}
%

\subsection{The tetraquark wave functions}
\label{sec:wavefunc}

To investigate the mass spectra of the tetraquark states, we need to construct the wave 
functions.
The total wave function is a direct product of the flavor, color, spin, and orbital wave 
functions.
For the $S$-wave states, the orbital wave function is symmetric.
%
Moreover, the Hamiltonian does not contain a flavor operator explicitly.
Thus we first construct the color-spin wave function, and then incorporate the flavor 
wave function to account for the Pauli principle.

The spins of the tetraquark states can be $0$, $1$ and $2$.
In the $qq{\otimes}\bar{q}\bar{q}$ configuration, the possible color-spin wave functions 
$\{\alpha_{i}^{J}\}$ are listed as follows,
\begin{enumerate}
\item $J^{P}=0^{+}$:
\begin{align}
&\alpha_{1}^{0}=\ket{\left(q_1q_2\right)_{1}^{6}\left(\bar{q}_{3}\bar{q}_{4}\right)_{1}^{\bar{6}}}_{0},
\notag\\
&\alpha_{2}^{0}=\ket{\left(q_1q_2\right)_{0}^{6}\left(\bar{q}_{3}\bar{q}_{4}\right)_{0}^{\bar{6}}}_{0},
\notag\\
&\alpha_{3}^{0}=\ket{\left(q_1q_2\right)_{1}^{\bar{3}}\left(\bar{q}_{3}\bar{q}_{4}\right)_{1}^{3}}_{0},
\notag\\
&\alpha_{4}^{0}=\ket{\left(q_1q_2\right)_{0}^{\bar{3}}\left(\bar{q}_{3}\bar{q}_{4}\right)_{0}^{3}}_{0},
\end{align}
\item $J^{P}=1^{+}$:
\begin{align}
&\alpha_{1}^{1}=\ket{\left(q_1q_2\right)_{1}^{6}\left(\bar{q}_{3}\bar{q}_{4}\right)_{1}^{\bar{6}}}_{1},
\notag\\
&\alpha_{2}^{1}=\ket{\left(q_1q_2\right)_{1}^{6}\left(\bar{q}_{3}\bar{q}_{4}\right)_{0}^{\bar{6}}}_{1},
\notag\\
&\alpha_{3}^{1}=\ket{\left(q_1q_2\right)_{0}^{6}\left(\bar{q}_{3}\bar{q}_{4}\right)_{1}^{\bar{6}}}_{1},
\notag\\
&\alpha_{4}^{1}=\ket{\left(q_1q_2\right)_{1}^{\bar{3}}\left(\bar{q}_{3}\bar{q}_{4}\right)_{1}^{3}}_{1},
\notag\\
&\alpha_{5}^{1}=\ket{\left(q_1q_2\right)_{1}^{\bar{3}}\left(\bar{q}_{3}\bar{q}_{4}\right)_{0}^{3}}_{1},
\notag\\
&\alpha_{6}^{1}=\ket{\left(q_1q_2\right)_{0}^{\bar{3}}\left(\bar{q}_{3}\bar{q}_{4}\right)_{1}^{3}}_{1},
\end{align}
\item $J^{P}=2^{+}$:
\begin{align}
&\alpha_{1}^{2}=\ket{\left(q_1q_2\right)_{1}^{6}\left(\bar{q}_{3}\bar{q}_{4}\right)_{1}^{\bar{6}}}_{2},
\notag\\
&\alpha_{2}^{2}=\ket{\left(q_1q_2\right)_{1}^{\bar{3}}\left(\bar{q}_{3}\bar{q}_{4}\right)_{1}^{3}}_{2},
\end{align}
\end{enumerate}
where the superscript $3$, $\bar{3}$, $6$, or $\bar{6}$ denotes the color, and the subscript 
$0$, $1$, or $2$ denotes the spin.

Next we consider the flavor wave function.
There are three types of total wave functions when we consider the Pauli principle:
%
\begin{enumerate}
\item Type A [$\text{Flavor}=\{cc\bar{c}\bar{c},bb\bar{b}\bar{b},cc\bar{b}\bar{b}\}$]:
\begin{enumerate}
\item $J^{P(C)}=0^{+(+)}$:
\begin{align}\label{eqn:wavefunc:total:A0}
\Psi_{A1}^{0^{+(+)}} = q_1q_2\bar{q}_3\bar{q}_4\otimes\alpha^{0}_{2},
\notag\\
\Psi_{A2}^{0^{+(+)}} = q_1q_2\bar{q}_3\bar{q}_4\otimes\alpha^{0}_{3},
\end{align}
\item $J^{P(C)}=1^{+(-)}$:
\begin{align}\label{eqn:wavefunc:total:A1}
\Psi_{A}^{1^{+(-)}} = q_1q_2\bar{q}_3\bar{q}_4\otimes\alpha^{1}_{4},
\end{align}
\item $J^{P(C)}=2^{+(+)}$:
\begin{align}\label{eqn:wavefunc:total:A2}
\Psi_{A}^{2^{+(+)}} = q_1q_2\bar{q}_3\bar{q}_4\otimes\alpha^{2}_{2},
\end{align}
\end{enumerate}
\item Type B [$\text{Flavor}=\{cc\bar{c}\bar{b},bb\bar{b}\bar{c}\}$]:
\begin{enumerate}
\item $J^{P}=0^{+}$:
\begin{align}\label{eqn:wavefunc:total:B0}
\Psi_{B1}^{0^{+}} = q_1q_2\bar{q}_3\bar{q}_4\otimes\alpha^{0}_{2},
\notag\\
\Psi_{B2}^{0^{+}} = q_1q_2\bar{q}_3\bar{q}_4\otimes\alpha^{0}_{3},
\end{align}
\item $J^{P}=1^{+}$:
\begin{align}\label{eqn:wavefunc:total:B1}
\Psi_{B1}^{1^{+}} = q_1q_2\bar{q}_3\bar{q}_4\otimes\alpha^{1}_{3},
\notag\\
\Psi_{B2}^{1^{+}} = q_1q_2\bar{q}_3\bar{q}_4\otimes\alpha^{1}_{4},
\notag\\
\Psi_{B3}^{1^{+}} = q_1q_2\bar{q}_3\bar{q}_4\otimes\alpha^{1}_{5},
\end{align}
\item $J^{P}=2^{+}$:
\begin{align}\label{eqn:wavefunc:total:B2}
\Psi_{B}^{2^{+}} = q_1q_2\bar{q}_3\bar{q}_4\otimes\alpha^{2}_{2},
\end{align}
\end{enumerate}
\item Type C [$\text{Flavor}=\{cb\bar{c}\bar{b}\}$]:
\begin{enumerate}
\item $J^{PC}=0^{++}$:
\begin{align}\label{eqn:wavefunc:total:C0}
\Psi_{C1}^{0^{++}} = q_1q_2\bar{q}_3\bar{q}_4\otimes\alpha^{0}_{1},
\notag\\
\Psi_{C2}^{0^{++}} = q_1q_2\bar{q}_3\bar{q}_4\otimes\alpha^{0}_{2},
\notag\\
\Psi_{C3}^{0^{++}} = q_1q_2\bar{q}_3\bar{q}_4\otimes\alpha^{0}_{3},
\notag\\
\Psi_{C4}^{0^{++}} = q_1q_2\bar{q}_3\bar{q}_4\otimes\alpha^{0}_{4},
\end{align}
\item $J^{PC}=1^{++}$:
\begin{align}\label{eqn:wavefunc:total:C1++}
&\Psi_{C1}^{1^{++}} = q_1q_2\bar{q}_3\bar{q}_4\otimes\frac{1}{\sqrt{2}}\left(\alpha^{1}_{2}+\alpha^{1}_{3}\right),
\notag\\
&\Psi_{C2}^{1^{++}} = q_1q_2\bar{q}_3\bar{q}_4\otimes\frac{1}{\sqrt{2}}\left(\alpha^{1}_{5}+\alpha^{1}_{6}\right),
\end{align}
\item $J^{PC}=1^{+-}$:
\begin{align}\label{eqn:wavefunc:total:C1+-}
&\Psi_{C1}^{1^{+-}} = q_1q_2\bar{q}_3\bar{q}_4\otimes\alpha^{1}_{1},
\notag\\
&\Psi_{C2}^{1^{+-}} = q_1q_2\bar{q}_3\bar{q}_4\otimes\frac{1}{\sqrt{2}}\left(\alpha^{1}_{2}-\alpha^{1}_{3}\right),
\notag\\
&\Psi_{C3}^{1^{+-}} = q_1q_2\bar{q}_3\bar{q}_4\otimes\alpha^{1}_{4},
\notag\\
&\Psi_{C4}^{1^{+-}} = q_1q_2\bar{q}_3\bar{q}_4\otimes\frac{1}{\sqrt{2}}\left(\alpha^{1}_{5}-\alpha^{1}_{6}\right),
\end{align}
\item $J^{PC}=2^{++}$:
\begin{align}\label{eqn:wavefunc:total:C2}
\Psi_{C1}^{2^{++}} = q_1q_2\bar{q}_3\bar{q}_4\otimes\alpha^{2}_{1},
\notag\\
\Psi_{C2}^{2^{++}} = q_1q_2\bar{q}_3\bar{q}_4\otimes\alpha^{2}_{2}.
\end{align}
\end{enumerate}
\end{enumerate}
Diagonalizing the Hamiltonian [Eq.~\eqref{eqn:hamiltonian:final}] in these bases, we can 
obtain the mass spectra and eigenvectors of the fully heavy tetraquarks.
%

\section{Numerical results}
\label{Sec:Result}

\subsection{Parameters}
\label{sec:Parameter}

\begin{table*}
	\centering
	\caption{Parameters of the $q\bar{q}$ and $qq$ pairs (in units of $\text{MeV}$).}
	\label{table:parameter}
	\begin{tabular}{lcccccccccccc}
		\toprule[1pt]
		\toprule[1pt]
		Parameter & $m_{n\bar{n}}$ & $m_{n\bar{s}}$ & $m_{s\bar{s}}$ & $m_{n\bar{c}}$ & $m_{s\bar{c}}$ & $m_{c\bar{c}}$ & $m_{n\bar{b}}$ & $m_{s\bar{b}}$ & $m_{c\bar{b}}$ & $m_{b\bar{b}}$ \\
		Value & $615.95$ & $794.22$ & $936.40$ & $1973.22$ & $2076.14$ & $3068.53$ & $5313.35$ & $5403.25$ & $6322.27$ & $9444.97$ \\
		Parameter & $v_{n\bar{n}}$ & $v_{n\bar{s}}$ & $v_{s\bar{s}}$ & $v_{n\bar{c}}$ & $v_{s\bar{c}}$ & $v_{c\bar{c}}$ & $v_{n\bar{b}}$ & $v_{s\bar{b}}$ & $v_{c\bar{b}}$ & $v_{b\bar{b}}$ \\
		Value & $477.92$ & $298.57$ & $249.18$ & $106.01$ & $107.87$ & $85.12$ & $33.89$ & $36.43$ & $47.18$ & $45.98$ \\
		Parameter & $m_{nn}$ & $m_{ns}$ &  $m_{ss}$ & $m_{nc}$ & $m_{sc}$ & $m_{cc}$ & $m_{nb}$ & $m_{sb}$ & $m_{cb}$ & $m_{b{b}}$ \\
		Value & $724.85$ & $906.65$ & $1049.36$ & $2079.96$ & $2183.68$ & $3171.51$ & $5412.25$ & $5494.80$ & $6416.07$ & $9529.57$ \\
		Parameter & $v_{n{n}}$  & $v_{n{s}}$ & $v_{ss}$ & $v_{n{c}}$ & $v_{s{c}}$ & $v_{c{c}}$ & $v_{n{b}}$ & $v_{s{b}}$ & $v_{c{b}}$ & $v_{b{b}}$  \\
		Value & $305.34$ & $212.75$ & $195.30$ & $62.81$ & $70.63$ & $56.75$ & $19.92$ & $8.47$ & $31.45$ & $30.65$ \\
		\bottomrule[1pt]
		\bottomrule[1pt]
	\end{tabular}
\end{table*}

To obtain the masses of fully heavy tetraquarks, we need to estimate the parameters. 
In Ref.~\cite{Weng:2018mmf}, we used the mesons to extract the parameters $m_{q\bar{q}}$ 
and $v_{q\bar{q}}$.
%
Then we used the light and singly heavy baryons to extract the $m_{qq}$ and $v_{qq}$ with 
at most one heavy quark.
Finally, we used a quark model consideration to estimate the $m_{QQ}$ and $v_{QQ}$.
The parameters are listed in Table~\ref{table:parameter}.
With these parameters, we can reproduce the meson and baryon masses.
%
Especially, we obtained the $\Xi_{cc}$ baryon at $3633.3\pm9.3~\text{MeV}$, which is 
very close to the LHCb's measurement~\cite{Aaij:2017ueg}.
In Ref.~\cite{Weng:2019ynv}, we further used these parameters to study the hidden-charm 
pentaquark states, and successfully reproduced the masses of the three newly observed 
$P_{c}$ states, $P_{c}(4312)$, $P_{c}(4440)$, and $P_{c}(4450)$~\cite{Aaij:2019vzc}, 
as well as the older one, $P_{c}(4380)$~\cite{Aaij:2015tga}.
In the present work, we use the same set of parameters to estimate the mass spectrum of 
the ground state $QQ\bar{Q}\bar{Q}$ tetraquarks.
%

\subsection{The $cc\bar{c}\bar{c}$ and $bb\bar{b}\bar{b}$ systems}
\label{sec:cccc+bbbb}

\begin{figure*}
\begin{tabular}{ccc}
\includegraphics[width=400pt]{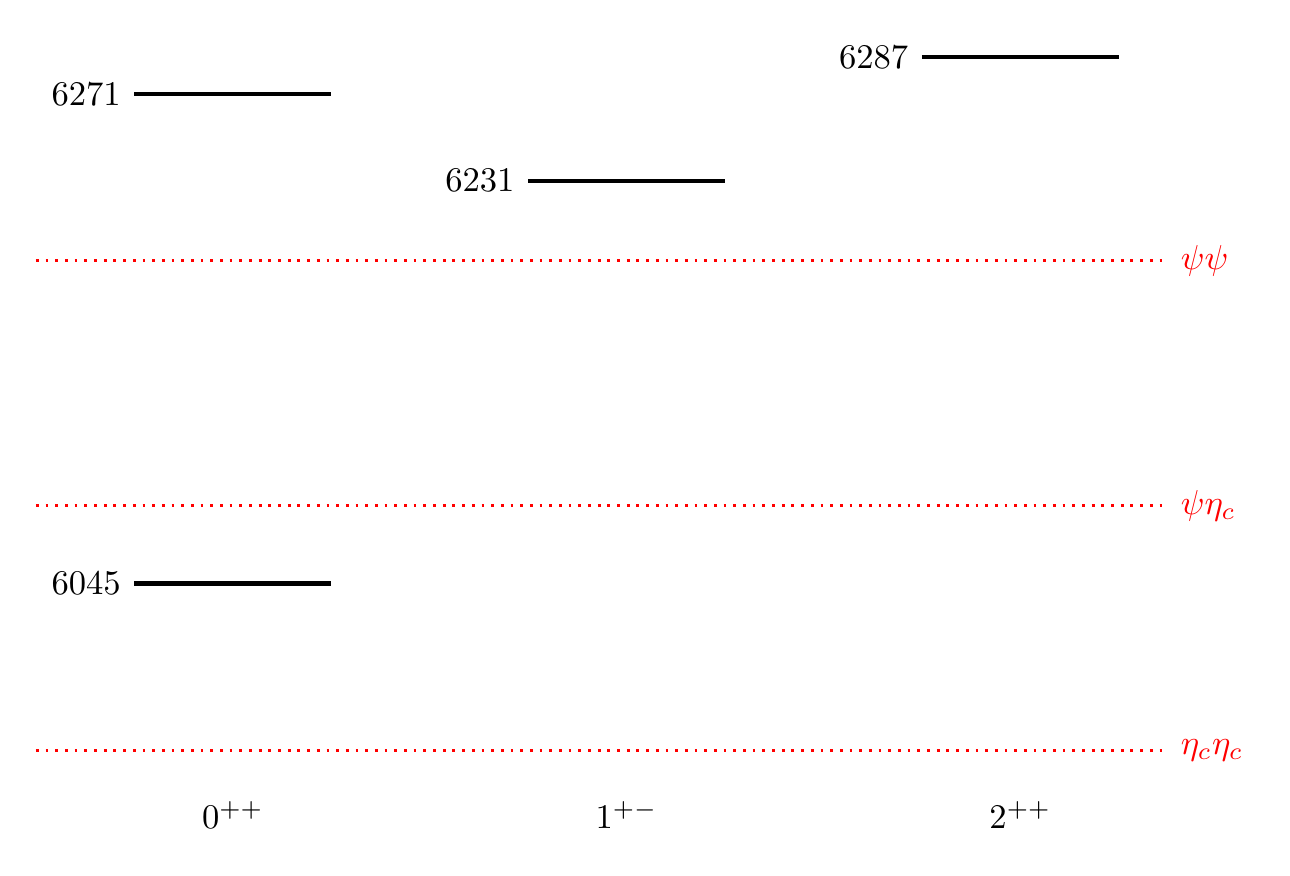}\\
(a) $cc\bar{c}\bar{c}$ states\\
&&\\
\includegraphics[width=400pt]{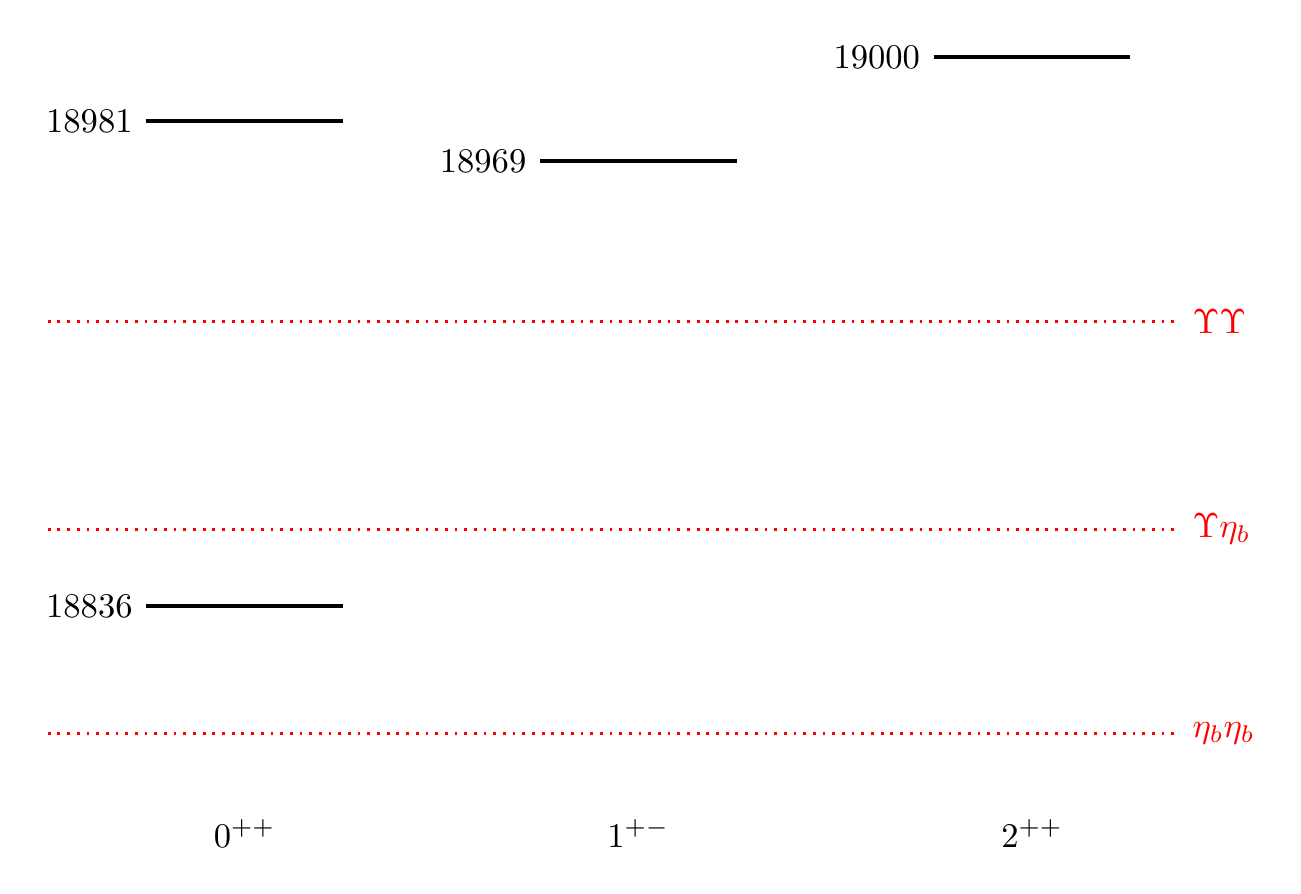}\\
(b) $bb\bar{b}\bar{b}$ states\\
\end{tabular}
\caption{Mass spectra of the $cc\bar{c}\bar{c}$ and $bb\bar{b}\bar{b}$ tetraquark states. The dotted lines indicate various meson-meson thresholds. The masses are all in units of MeV.}
\label{fig:cccc+bbbb}
\end{figure*}
%
\begin{table}
	\centering
	\caption{Masses and eigenvectors of the $cc\bar{c}\bar{c}$ and $bb\bar{b}\bar{b}$ tetraquarks. The masses are all in units of MeV.}
	\label{table:mass:cccc+bbbb}
	\begin{tabular}{ccccccc}
		\toprule[1pt]
		\toprule[1pt]
		System & $J^{PC}$ & Mass & Eigenvector \\
		\midrule[1pt]
		$cc\bar{c}\bar{c}$ & $0^{++}$ & $6044.9$ & $\{0.834,0.552\}$ \\
		&& $6271.3$ & $\{-0.552,0.834\}$ \\
		& $1^{+-}$ & $6230.6$ & $\{1\}$ \\
		& $2^{++}$ & $6287.3$ & $\{1\}$ \\
		\midrule[1pt]
		$bb\bar{b}\bar{b}$ & $0^{++}$ & $18836.1$ & $\{0.903,0.431\}$ \\
		&& $18981.0$ & $\{-0.431,0.903\}$ \\
		& $1^{+-}$ & $18969.4$ & $\{1\}$ \\
		& $2^{++}$ & $19000.1$ & $\{1\}$ \\
		\bottomrule[1pt]
		\bottomrule[1pt]
	\end{tabular}
\end{table}
%
\begin{table}
	\centering
	\caption{The eigenvectors of the $cc\bar{c}\bar{c}$ tetraquarks in the $c\bar{c}{\otimes}c\bar{c}$ configuration. The masses are all in units of MeV.}
	\label{table:wavefunc:cccc:13x24}
	\begin{tabular}{ccccccc}
		\toprule[1pt]
		\toprule[1pt]
		System & $J^{PC}$ & Mass & $\psi\psi$ & $\psi\eta_{c}$ & $\eta_{c}\psi$ & $\eta_{c}\eta_{c}$ \\
		\midrule[1pt]
		$cc\bar{c}\bar{c}$ & $0^{++}$ & $6044.9$ & $0.430$ &&& $0.616$ \\
		&& $6271.3$ & $-0.631$ &&& $0.191$ \\
		& $1^{+-}$ & $6230.6$ && $0.408$ & $0.408$ \\
		& $2^{++}$ & $6287.3$ & $0.577$ \\
		\bottomrule[1pt]
		\bottomrule[1pt]
	\end{tabular}
\end{table}
%
\begin{table}
	\centering
	\caption{The eigenvectors of the $bb\bar{b}\bar{b}$ tetraquarks in the $b\bar{b}{\otimes}b\bar{b}$ configuration. The masses are all in units of MeV.}
	\label{table:wavefunc:bbbb:13x24}
	\begin{tabular}{ccccccc}
		\toprule[1pt]
		\toprule[1pt]
		System & $J^{PC}$ & Mass & $\Upsilon\Upsilon$ & $\Upsilon\eta_{b}$ & $\eta_{b}\Upsilon$ & $\eta_{b}\eta_{b}$ \\
		\midrule[1pt]
		$bb\bar{b}\bar{b}$ & $0^{++}$ & $18836.1$ & $0.514$ &&& $0.584$ \\
		&& $18981.0$ & $-0.565$ &&& $0.275$ \\
		& $1^{+-}$ & $18969.4$ && $0.408$ & $0.408$ \\
		& $2^{++}$ & $19000.1$ & $0.577$ \\
		\bottomrule[1pt]
		\bottomrule[1pt]
	\end{tabular}
\end{table}
%
\begin{table}
	\centering
	\caption{The values of $k\cdot|c_{i}|^2$ for the $cc\bar{c}\bar{c}$ tetraquarks (in units of MeV).}
	\label{table:kc_i^2:cc.cc}
	\begin{tabular}{ccccccccccccc}
		\toprule[1pt]
		\toprule[1pt]
		System & $J^{PC}$ & Mass & $\psi\psi$ & $\psi\eta_{c}$ & $\eta_{c}\eta_{c}$ \\
		\midrule[1pt]
		$cc\bar{c}\bar{c}$&$0^{++}$&6044.9&0&&182.8 \\
		&&6271.3&195.8&&35.3 \\
		&$1^{+-}$&6230.6&&226.3 \\
		&$2^{++}$&6287.3&180.1 \\
		\bottomrule[1pt]
		\bottomrule[1pt]
	\end{tabular}
\end{table}
%
\begin{table}
	\centering
	\caption{The values of $k\cdot|c_{i}|^2$ for the $bb\bar{b}\bar{b}$ tetraquarks (in units of MeV).}
	\label{table:kc_i^2:bb.bb}
	\begin{tabular}{ccccccccccccc}
		\toprule[1pt]
		\toprule[1pt]
		System & $J^{PC}$ & Mass & $\Upsilon\Upsilon$ & $\Upsilon\eta_{b}$ & $\eta_{b}\eta_{b}$ \\
		\midrule[1pt]
		$bb\bar{b}\bar{b}$&$0^{++}$&18836.1&0&&205.6 \\
		&&18981.0&241.5&&99.9 \\
		&$1^{+-}$&18969.4&&340.6 \\
		&$2^{++}$&19000.1&289.4 \\
		\bottomrule[1pt]
		\bottomrule[1pt]
	\end{tabular}
\end{table}
%
\begin{table}
	\centering
	\caption{The partial width ratios for the $cc\bar{c}\bar{c}$ tetraquarks. For each state, we chose one mode as the reference channel, and the partial width ratios of the other channels are calculated relative to this channel. The masses are all in units of MeV.}
	\label{table:R:cc.cc}
	\begin{tabular}{ccccccccccccc}
		\toprule[1pt]
		\toprule[1pt]
		System & $J^{PC}$ & Mass & $\psi\psi$ & $\psi\eta_{c}$ & $\eta_{c}\eta_{c}$ \\
		\midrule[1pt]
		$cc\bar{c}\bar{c}$&$0^{++}$&6044.9&0&&1 \\
		&&6271.3&5.6&&1 \\
		&$1^{+-}$&6230.6&&1 \\
		&$2^{++}$&6287.3&1 \\
		\bottomrule[1pt]
		\bottomrule[1pt]
	\end{tabular}
\end{table}
%
\begin{table}
	\centering
	\caption{The partial width ratios for the $bb\bar{b}\bar{b}$ tetraquarks. For each state, we chose one mode as the reference channel, and the partial width ratios of the other channels are calculated relative to this channel. The masses are all in units of MeV.}
	\label{table:R:bb.bb}
	\begin{tabular}{ccccccccccccc}
		\toprule[1pt]
		\toprule[1pt]
		System & $J^{PC}$ & Mass & $\Upsilon\Upsilon$ & $\Upsilon\eta_{b}$ & $\eta_{b}\eta_{b}$ \\
		\midrule[1pt]
		$bb\bar{b}\bar{b}$&$0^{++}$&18836.1&0&&1 \\
		&&18981.0&2.4&&1 \\
		&$1^{+-}$&18969.4&&1 \\
		&$2^{++}$&19000.1&1 \\
		\bottomrule[1pt]
		\bottomrule[1pt]
	\end{tabular}
\end{table}

Inserting the parameters into the Hamiltonian, we can obtain the mass spectra of tetraquarks.
The masses and eigenvectors of the $cc\bar{c}\bar{c}$ and $bb\bar{b}\bar{b}$ tetraquarks 
are listed in Table~\ref{table:mass:cccc+bbbb}.
In the following, we will use $T(QQ\bar{Q}\bar{Q},m,J^{PC})$ to denote the 
$QQ\bar{Q}\bar{Q}$ tetraquarks.
In both cases, the lightest states have quantum number $J^{PC}=0^{++}$.
They are $T(cc\bar{c}\bar{c},6044.9,0^{++})$ and $T(bb\bar{b}\bar{b},18836.1,0^{++})$, 
respectively.
In Fig.~\ref{fig:cccc+bbbb}, we plot the relative position of the $cc\bar{c}\bar{c}$ and 
$bb\bar{b}\bar{b}$ tetraquarks, along with meson-meson thresholds which they can decay 
into through quark rearrangement.
From the figure, we can easily see that all states are above thresholds.
Among them, the two ground states are only above the thresholds of the two pseudoscalar 
mesons, while the other states are all above the thresholds of two vector mesons.

Besides the masses, the eigenvectors also provide important information of the tetraquarks.
The $0^{++}$ states are of particular interests since they have two bases.
Their color configurations are $\ket{(QQ)^{6_{c}}{\otimes}(\bar{Q}\bar{Q})^{\bar{6}_{c}}}$ and $\ket{(QQ)^{\bar{3}_{c}}{\otimes}(\bar{Q}\bar{Q})^{{3}_{c}}}$ respectively.
For simplicity, we denote them as $6_{c}{\otimes}\bar{6}_{c}$ and $\bar{3}_{c}{\otimes}{3}_{c}$.
In the one-gluon-exchange (OGE) picture, the interaction between two quarks are attractive if they combine into a $\bar{3}_{c}$ diquark $(qq)^{\bar{3}_{c}}$, while repulsive if they combine into a ${6}_{c}$ diquark $(qq)^{{6}_{c}}$.
However, the attraction between a $6_{c}$ diquark and a $\bar{6}_{c}$ anti-diquark is much stronger than that between the $\bar{3}_{c}{\otimes}{3}_{c}$ counterpart.
The two competing effects make the tetraquark much more complicated compared to the ordinary hadrons~\cite{Wang:2019rdo}.
In Table~\ref{table:mass:cccc+bbbb}, we present the eigenvectors of the tetraquark states.
We see that the ground states are both dominated by the $6_{c}{\otimes}\bar{6}_{c}$ components.
More precisely, the $T(cc\bar{c}\bar{c},6044.9,0^{++})$ state has $69.5\%$ of the $6_{c}{\otimes}\bar{6}_{c}$ component and the $T(bb\bar{b}\bar{b},18836.1,0^{++})$ state has $81.5\%$.
Here we illustrate the underlying dynamics as follows.
We first consider the color interaction.
Because of the symmetry between quarks (or antiquarks), the interaction strength between two quarks equals to that between two antiquarks, and the $Q\bar{Q}$ interactions share one strength.
More precisely, we have (taking $bb\bar{b}\bar{b}$ as an example)
\begin{align}\label{eqn:HCE:bbbb}
&
\Braket{H_{\text{C}}\left(bb\bar{b}\bar{b}\right)}
\notag\\
={}&
-\frac{3}{4}
\Braket{
m_{bb}
\left(V^{\text{C}}_{12}+V^{\text{C}}_{34}\right)
+
m_{b\bar{b}}
\left(V^{\text{C}}_{13}+V^{\text{C}}_{24}+V^{\text{C}}_{14}+V^{\text{C}}_{23}\right)}
\notag\\
={}&
-\frac{3}{4}
\Braket{
m_{b\bar{b}}
\sum_{i<j}
V^{\text{C}}_{ij}
+
2{\delta}m_{b}
\left(V^{\text{C}}_{12}+V^{\text{C}}_{34}\right)}
\notag\\
={}&
2m_{b\bar{b}}
-\frac{3}{2}
{\delta}m_{b}
\Braket{V^{\text{C}}_{12}+V^{\text{C}}_{34}}
\notag\\
={}&
2m_{b\bar{b}}
+
{\delta}m_{b}
\begin{pmatrix}
-1&0\\
0&+2
\end{pmatrix}
\end{align}
where ${\delta}m_{b}{\equiv}(m_{bb}-m_{b\bar{b}})/2$.
%
%
%
%
%
In Ref.~\cite{Weng:2018mmf}, we have estimated the ${\delta}m_{b}$ to be $42.30~\text{MeV}$.
Here we see that the color interaction do not mix the $6_{c}{\otimes}\bar{6}_{c}$ and $\bar{3}_{c}{\otimes}3_{c}$ configurations.
Moreover, the color interaction favors the $6_{c}{\otimes}\bar{6}_{c}$ configuration over the $\bar{3}_{c}{\otimes}3_{c}$ one.
%
%
Note that these conclusions still hold if we include the internal dynamics in the quark model~\cite{Wang:2019rdo,Deng:2020iqw}.
Similarly, the CM interaction in the two bases is
\begin{align}
\Braket{H_{\text{CM}}\left(bb\bar{b}\bar{b}\right)}
={}&
\begin{pmatrix}
\frac{1}{2}v_{bb}&\sqrt{\frac{3}{2}}v_{b\bar{b}}\\
\sqrt{\frac{3}{2}}v_{b\bar{b}}&\frac{1}{3}v_{bb}-\frac{2}{3}v_{b\bar{b}}
\end{pmatrix}
\notag\\
={}&
v_{b\bar{b}}
\begin{pmatrix}
\frac{1}{3}&\sqrt{\frac{3}{2}}\\
\sqrt{\frac{3}{2}}&-\frac{4}{9}
\end{pmatrix}\,.
\end{align}
In the last line, we have used 
$v_{bb}/v_{b\bar{b}}{=}2/3$, 
which was estimated from the meson and baryon spectra~\cite{Weng:2018mmf} (see Table~\ref{table:parameter}) and also interpreted in the quark model using the Cornell or Logarithmic potential~\cite{Keren-Zur:2008:ukn01}.
Here, $v_{b\bar{b}}=45.98~\text{MeV}$.
If we ignore the off-diagonal terms, we see that the CM interaction favors the $\bar{3}_{c}{\otimes}3_{c}$ configuration, which is opposite to that of colorelectric interaction.
However, the splitting induced by the CM interaction is much smaller than that of the colorelectric interaction ($\sim28\%$), thus the net effect is that the $6_{c}{\otimes}\bar{6}_{c}$ configuration has lower mass.
Similar pattern exists in the bound state of quantum electrodynamics, where the Coulomb interaction determines the energy bands, and the hyperfine interaction gives small shift of each band.
%
Including the off-diagonal terms will push down the lower eigenstate and raise the upper eigenstate, thus the proceeding conclusion still holds. 
If we go to the $cc\bar{c}\bar{c}$ system, the splitting induced by the colorelectric interaction does not change much, while the CM interaction becomes larger ($v_{q\bar{q}}\sim1/m_{q}m_{\bar{q}}$).
%
The net effect is that the relative size between the off-diagonal terms and the splitting induced by diagonal terms becomes larger, which causes much stronger mixing in the $cc\bar{c}\bar{c}$ system than in the $bb\bar{b}\bar{b}$ system, as shown in Table~\ref{table:mass:cccc+bbbb}.
%
%

Next we consider the decay properties of the $cc\bar{c}\bar{c}$ and $bb\bar{b}\bar{b}$ tetraquarks.
%
%
%
%
In the $Q\bar{Q}{\otimes}Q\bar{Q}$ configuration, the color wave function of the tetraquark falls into two categories:
the color-singlet $\ket{(Q\bar{Q})^{1_{c}}(Q\bar{Q})^{1_{c}}}$ and 
the color-octet $\ket{(Q\bar{Q})^{8_{c}}(Q\bar{Q})^{8_{c}}}$.
%
%
The former one can easily decay into two $S$-wave mesons (the so-called ``Okubo-Zweig-Iizuka- (OZI-)superallowed'' decays), and the latter one can only fall apart through the gluon exchange~\cite{Jaffe:1976ig,Strottman:1979qu}.
%
In this work, we will focus on the ``OZI-superallowed'' decays.
We transform the eigenvectors of the tetraquark states into the $c\bar{c}{\otimes}c\bar{c}$/$b\bar{b}{\otimes}b\bar{b}$ configuration.
The corresponding eigenvectors are listed in Tables~\ref{table:wavefunc:cccc:13x24}--\ref{table:wavefunc:bbbb:13x24}.
For simplicity, we only present the color-singlet components, and we rewrite the bases as a direct product of two mesons.
For each decay mode, the branching fraction is proportional to the square of the coefficient $c_i$ of the corresponding component in the eigenvectors, and also depends on the phase space.
For two body decay through $L$-wave, the partial decay width reads~\cite{Gao-1992-Group,Weng:2019ynv}
\begin{equation}\label{eqn:width}
\Gamma_{i}=\gamma_{i}\alpha\frac{k^{2L+1}}{m^{2L}}{\cdot}|c_i|^2,
\end{equation}
where $\gamma_{i}$ is a quantity determined by the decay dynamics, $\alpha$ is an effective coupling constant, $m$ is the mass of the initial state, and $k$ is the momentum of the final states in the rest frame of the initial state.
For the decays of the $S$-wave tetraquarks, $(k/m)^2$ is of order $10^{-2}$ or even smaller.
All higher wave decays are suppressed.
%
Thus we will only consider the $S$-wave decays in this work. 
Employing the eigenvectors in Tables~\ref{table:wavefunc:cccc:13x24}--\ref{table:wavefunc:bbbb:13x24}, we can calculate the value of $k\cdot|c_i|^2$ for each decay process (see Tables~\ref{table:kc_i^2:cc.cc}--\ref{table:kc_i^2:bb.bb}).
Next we should consider the $\gamma_{i}$.
Generally, $\gamma_{i}$ is determined by the spatial wave functions of both initial and final states, which are different for each decay process.
In the quark model, the spatial wave functions of the pseudoscalar and vector mesons are the same.
Thus for each tetraquark, we have
\begin{equation}
\gamma_{{\psi}{\psi}}
=
\gamma_{{\psi}{\eta_{c}}}
=
\gamma_{{\eta_{c}}{\eta_{c}}}
\end{equation}
and
\begin{equation}
\gamma_{{\Upsilon}{\Upsilon}}
=
\gamma_{{\Upsilon}{\eta_{b}}}
=
\gamma_{{\eta_{b}}{\eta_{b}}}\,.
\end{equation}
The values of the relative widths of different decay processes are listed in Tables~\ref{table:R:cc.cc} and~\ref{table:R:bb.bb}.
For the $cc\bar{c}\bar{c}$ tetraquarks, the ground state of the $0^{++}$ is above the dissociation channel $\eta_{c}\eta_{c}$, thus it may be \emph{broad}~\cite{Jaffe:1976ig}.
The other scalar state, $T(cc\bar{c}\bar{c},6271.3,0^{++})$, has two $S$-wave decay modes, namely $\psi\psi$ and $\eta_{c}\eta_{c}$, their relative decay width ratio is
\begin{equation}
\frac{\Gamma\left[T(cc\bar{c}\bar{c},6271.3,0^{++}){\to}\psi\psi\right]}{\Gamma\left[T(cc\bar{c}\bar{c},6271.3,0^{++}){\to}\eta_{c}\eta_{c}\right]}
=
5.6\,.
\end{equation}
The dominant decay mode is the $\psi\psi$ final state.
The $bb\bar{b}\bar{b}$ tetraquarks are very similar to the $cc\bar{c}\bar{c}$ tetraquarks.
The ground state $T(bb\bar{b}\bar{b},18836.1,0^{++})$ can only decay into $\eta_{b}\eta_{b}$, while the higher scalar state $T(bb\bar{b}\bar{b},18981.0,0^{++})$ can decay into both $\Upsilon\Upsilon$ and $\eta_{b}\eta_{b}$ channels, with relative decay with ratio
\begin{equation}
\frac{\Gamma\left[T(bb\bar{b}\bar{b},18981.0,0^{++}){\to}\Upsilon\Upsilon\right]}{\Gamma\left[T(bb\bar{b}\bar{b},18981.0,0^{++}){\to}\eta_{b}\eta_{b}\right]}
=
2.4\,.
\end{equation}
Thus the widths of the two modes do no differ very much.
%

\subsection{The $cc\bar{c}\bar{b}$ and $bb\bar{b}\bar{c}$ systems}
\label{sec:cccb+bbbc}

\begin{figure*}
	\begin{tabular}{ccc}
		\includegraphics[width=400pt]{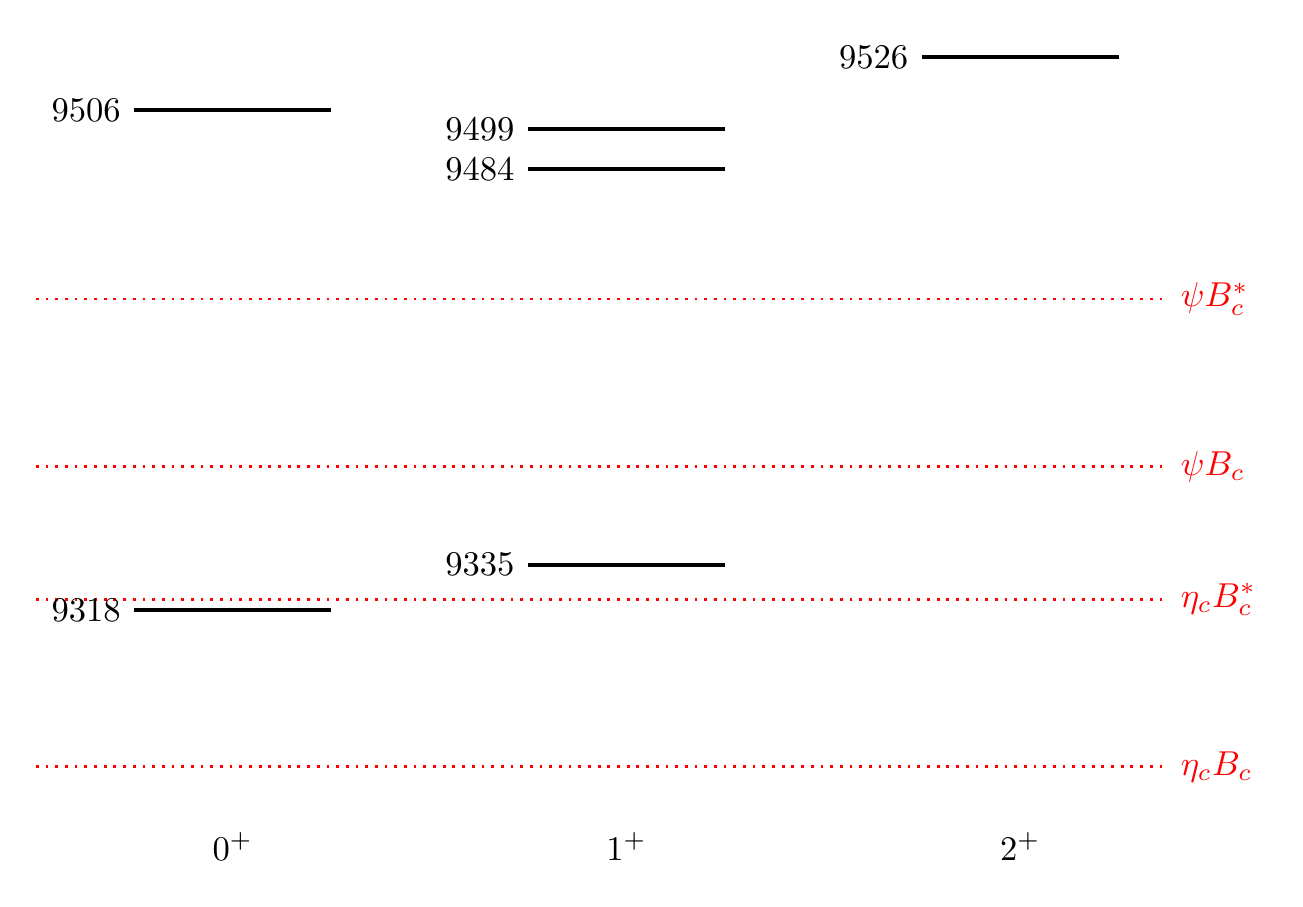}\\
		(a) $cc\bar{c}\bar{b}$ states\\
		&&\\
		\includegraphics[width=400pt]{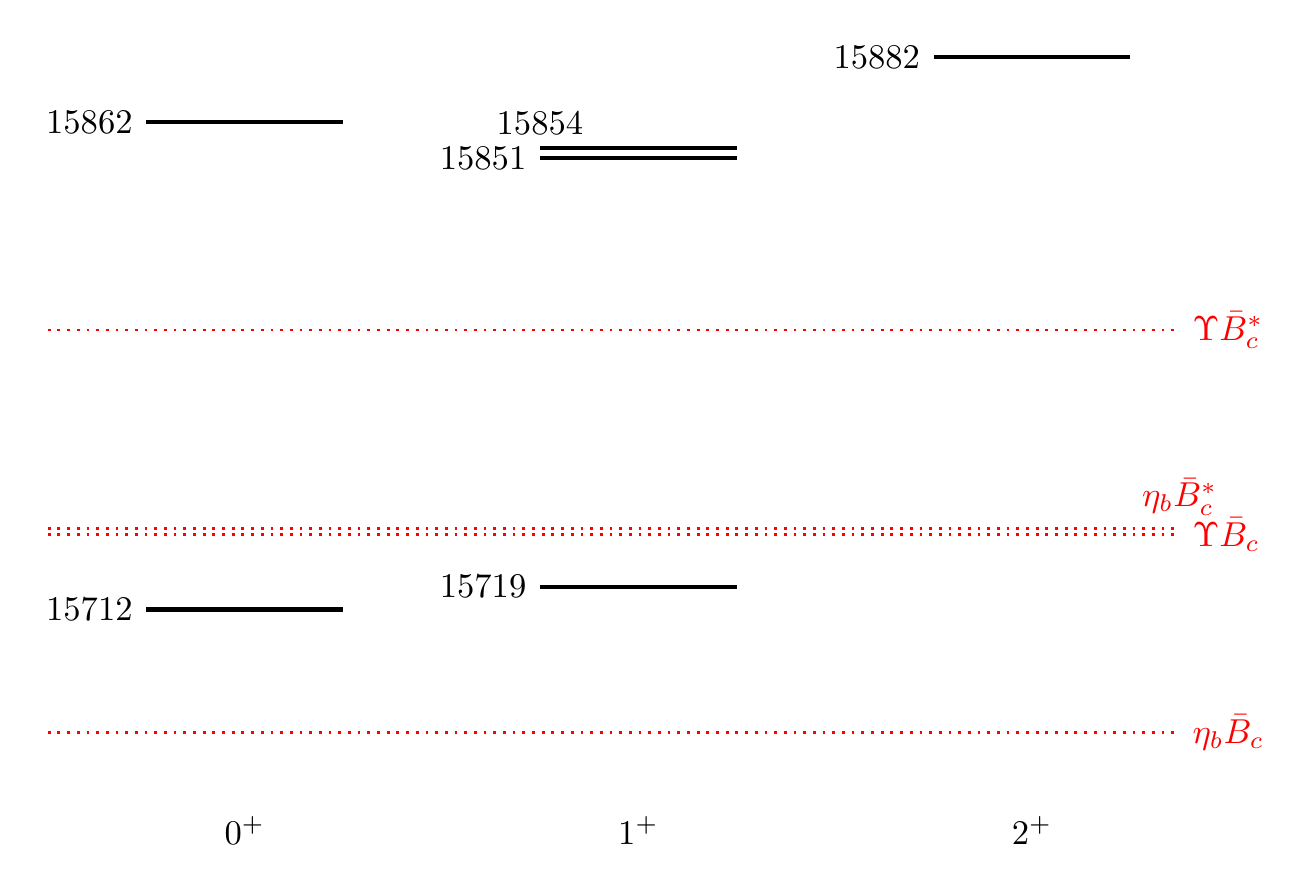}\\
		(b) $bb\bar{b}\bar{c}$ states\\
	\end{tabular}
	\caption{Mass spectra of the $cc\bar{c}\bar{b}$ and $bb\bar{b}\bar{c}$ tetraquark states. The dotted lines indicate various meson-meson thresholds. Here the predicted mass $M_{B_{c}^{*}}=6338~\text{MeV}$ of Godfrey {\it et~al.}~\cite{Godfrey:1985xj} is used. The masses are all in units of MeV.}
	\label{fig:cccb+bbbc}
\end{figure*}
%
\begin{table}
	\centering
	\caption{Masses and eigenvectors of the $cc\bar{c}\bar{b}$ and $bb\bar{b}\bar{c}$ tetraquarks. The masses are all in units of MeV.}
	\label{table:mass:cccb+bbbc}
	\begin{tabular}{ccccccc}
		\toprule[1pt]
		\toprule[1pt]
		System&$J^{P}$&Mass&Eigenvector \\
		\midrule[1pt]
		$cc\bar{c}\bar{b}$&$0^{+}$&9317.5&$\{0.869,0.495\}$ \\
		&&9505.9&$\{-0.495,0.869\}$ \\
		&$1^{+}$&9335.1&$\{0.941,0.140,-0.306\}$ \\
		&&9484.3&$\{0.190,-0.972,0.138\}$ \\
		&&9498.5&$\{0.279,0.188,0.942\}$ \\
		&$2^{+}$&9525.9&$\{1\}$ \\
		\midrule[1pt]
		$bb\bar{b}\bar{c}$&$0^{+}$&15711.9&$\{0.908,0.418\}$ \\
		&&15862.0&$\{-0.418,0.908\}$ \\
		&$1^{+}$&15719.1&$\{0.968,-0.005,-0.252\}$ \\
		&&15851.3&$\{-0.015,-0.999,-0.037\}$ \\
		&&15854.4&$\{-0.251,0.039,-0.967\}$ \\
		&$2^{+}$&15882.3&$\{1\}$ \\
		\bottomrule[1pt]
		\bottomrule[1pt]
	\end{tabular}
\end{table}
%
\begin{table}
	\centering
	\caption{The eigenvectors of the $cc\bar{c}\bar{b}$ tetraquarks in the $c\bar{c}{\otimes}c\bar{b}$ configuration. The masses are all in units of MeV.}
	\label{table:wavefunc:cccb:13x24}
	\begin{tabular}{ccccccc}
		\toprule[1pt]
		\toprule[1pt]
		System&$J^{P}$&Mass&${\psi}{B}_{c}^{*}$&${\psi}{B}_{c}$&${\eta}_{c}{B}_{c}^{*}$&${\eta}_{c}{B}_{c}^{*}$\\
		\midrule[1pt]
		$cc\bar{c}\bar{b}$&$0^{+}$&9317.5&0.472&&&0.602\\
		&&9505.9&$-0.601$&&&0.232\\
		&$1^{+}$&9335.1&0.418&$-0.416$&0.530\\
		&&9484.3&0.166&$-0.435$&$-0.359$\\
		&&9498.5&$-0.545$&$-0.235$&0.081\\
		&$2^{+}$&9525.9&0.577\\
		\bottomrule[1pt]
		\bottomrule[1pt]
	\end{tabular}
\end{table}
%
\begin{table}
	\centering
	\caption{The eigenvectors of the $bb\bar{b}\bar{c}$ tetraquarks in the $b\bar{b}{\otimes}b\bar{c}$ configuration. The masses are all in units of MeV.}
	\label{table:wavefunc:bbbc:13x24}
	\begin{tabular}{ccccccc}
		\toprule[1pt]
		\toprule[1pt]
		System&$J^{P}$&Mass&${\Upsilon}\bar{B}_{c}^{*}$&${\Upsilon}\bar{B}_{c}$&$\eta_{b}\bar{B}_{c}^{*}$&$\eta_{b}\bar{B}_{c}$\\
		\midrule[1pt]
		$bb\bar{b}\bar{c}$&$0^{+}$&15711.9&0.521&&&0.580\\
		&&15862.0&$-0.558$&&&0.283\\
		&$1^{+}$&15719.1&0.456&$-0.470$&0.466\\
		&&15851.3&$-0.023$&$-0.413$&$-0.403$\\
		&&15854.4&$-0.540$&$-0.161$&$0.193$\\
		&$2^{+}$&15882.3&0.577\\
		\bottomrule[1pt]
		\bottomrule[1pt]
	\end{tabular}
\end{table}
%
\begin{table}
	\centering
	\caption{The values of $k\cdot|c_{i}|^2$ for the $cc\bar{c}\bar{b}$ tetraquarks (in units of MeV).}
	\label{table:kc_i^2:cc.cb}
	\begin{tabular}{ccccccccccccc}
		\toprule[1pt]
		\toprule[1pt]
		System&$J^{P}$&Mass&${\psi}{B}_{c}^{*}$&${\psi}{B}_{c}$&${\eta}_{c}{B}_{c}^{*}$&${\eta}_{c}{B}_{c}^{*}$\\
		\midrule[1pt]
		$cc\bar{c}\bar{b}$&$0^{+}$&9317.5&$\times$&&&177.2\\
		&&9505.9&196.8&&&54.9\\
		&$1^{+}$&9335.1&$\times$&$\times$&65.1\\
		&&9484.3&12.5&129.6&105.4\\
		&&9498.5&153.4&40.2&5.7\\
		&$2^{+}$&9525.9&205.9\\
		\bottomrule[1pt]
		\bottomrule[1pt]
	\end{tabular}
\end{table}
%
\begin{table}
	\centering
	\caption{The values of $k\cdot|c_{i}|^2$ for the $bb\bar{b}\bar{c}$ tetraquarks (in units of MeV).}
	\label{table:kc_i^2:bb.bc}
	\begin{tabular}{ccccccccccccc}
		\toprule[1pt]
		\toprule[1pt]
		System&$J^{P}$&Mass&${\Upsilon}\bar{B}_{c}^{*}$&${\Upsilon}\bar{B}_{c}$&$\eta_{b}\bar{B}_{c}^{*}$&$\eta_{b}\bar{B}_{c}$\\
		\midrule[1pt]
		$bb\bar{b}\bar{c}$&$0^{+}$&15711.9&$\times$&&&180.7\\
		&&15862.0&216.8&&&95.9\\
		&$1^{+}$&15719.1&$\times$&$\times$&$\times$\\
		&&15851.3&0.3&159.6&151.8\\
		&&15854.4&190.4&24.5&35.1\\
		&$2^{+}$&15882.3&266.6\\
		\bottomrule[1pt]
		\bottomrule[1pt]
	\end{tabular}
\end{table}
%
\begin{table}
	\centering
	\caption{The partial width ratios for the $cc\bar{c}\bar{b}$ tetraquarks. For each state, we chose one mode as the reference channel, and the partial width ratios of the other channels are calculated relative to this channel. The masses are all in units of MeV.}
	\label{table:R:cc.cb}
	\begin{tabular}{ccccccccccccc}
		\toprule[1pt]
		\toprule[1pt]
		System&$J^{P}$&Mass&${\psi}{B}_{c}^{*}$&${\psi}{B}_{c}$&${\eta}_{c}{B}_{c}^{*}$&${\eta}_{c}{B}_{c}^{*}$\\
		\midrule[1pt]
		$cc\bar{c}\bar{b}$&$0^{+}$&9317.5&$\times$&&&1\\
		&&9505.9&3.6&&&1\\
		&$1^{+}$&9335.1&$\times$&$\times$&1\\
		&&9484.3&1&10.4&8.4\\
		&&9498.5&1&0.3&0.04\\
		&$2^{+}$&9525.9&1\\
		\bottomrule[1pt]
		\bottomrule[1pt]
	\end{tabular}
\end{table}
%
\begin{table}
	\centering
	\caption{The partial width ratios for the $bb\bar{b}\bar{c}$ tetraquarks. For each state, we chose one mode as the reference channel, and the partial width ratios of the other channels are calculated relative to this channel. The masses are all in units of MeV.}
	\label{table:R:bb.bc}
	\begin{tabular}{ccccccccccccc}
		\toprule[1pt]
		\toprule[1pt]
		System&$J^{P}$&Mass&${\Upsilon}\bar{B}_{c}^{*}$&${\Upsilon}\bar{B}_{c}$&$\eta_{b}\bar{B}_{c}^{*}$&$\eta_{b}\bar{B}_{c}$\\
		\midrule[1pt]
		$bb\bar{b}\bar{c}$&$0^{+}$&15711.9&$\times$&&&1\\
		&&15862.0&2.3&&&1\\
		&$1^{+}$&15719.1&$\times$&$\times$&$\times$\\
		&&15851.3&0.002&1.1&1\\
		&&15854.4&5.4&0.7&1\\
		&$2^{+}$&15882.3&1\\
		\bottomrule[1pt]
		\bottomrule[1pt]
	\end{tabular}
\end{table}

Next we consider the $cc\bar{c}\bar{b}$ and $bb\bar{b}\bar{c}$ tetraquarks.
%
In Table~\ref{table:mass:cccb+bbbc}, we list the masses and eigenvectors of these states.
Their relative position and possible decay channels are plotted in Fig.~\ref{fig:cccb+bbbc}.
Since the $B_{c}^{*}$ has not been observed in experiment~\cite{Zyla:2020zbs}, we use the $M_{B_{c}^{*}}=6338~\text{MeV}$ predicted by the Godfrey-Isgur model~\cite{Godfrey:1985xj} to estimate the dimeson thresholds.
%

For these states, the two antiquarks do not have to obey the Pauli principle.
%
We have three, rather than one, bases with $J^{P}=1^{+}$, which provides us a new platform to to study the color mixing.
First we consider the $cc\bar{c}\bar{b}$ system.
The ground state is $T(cc\bar{c}\bar{b},9317.5,0^{+})$.
This state is dominated by the color-sextet component ($75.5\%$), just as the the $QQ\bar{Q}'\bar{Q}'$ tetraquarks.
The reason is similar to that in the $QQ\bar{Q}'\bar{Q}'$ tetraquarks.
It is interesting that the same mechanism can also apply to the $1^{+}$ states.
For the $cc\bar{c}\bar{b}$ tetraquarks, the color interaction reads
\begin{align}\label{eqn:HCE:cccb}
\Braket{H_{\text{C}}\left(cc\bar{c}\bar{b}\right)}
=
m_{c\bar{c}}
+
m_{c\bar{b}}
-
\frac{3}{2}
{\delta}m'
\Braket{V^{\text{C}}_{12}+V^{\text{C}}_{34}}
\end{align}
where ${\delta}m'=(3{\delta}m_{b}+{\delta}m_{c})/4=42.89~\text{MeV}$.
There are three bases with $J^{P}=1^{+}$, namely $\ket{(cc)_{0}^{6}(\bar{c}\bar{b})_{1}^{\bar{6}}}$, $\ket{(cc)_{1}^{\bar{3}}(\bar{c}\bar{b})_{1}^{3}}$ and $\ket{(cc)_{1}^{\bar{3}}(\bar{c}\bar{b})_{0}^{3}}$ [see Eq.~\eqref{eqn:wavefunc:total:B1}].
Inserting these bases into the color interaction [Eq.~\eqref{eqn:HCE:cccb}], we have
\begin{align}\label{eqn:HCE:cccb:J1}
\Braket{H_{\text{C}}\left(cc\bar{c}\bar{b}\right)}
=
m_{c\bar{c}}
+
m_{c\bar{b}}
+
{\delta}m'
\begin{pmatrix}
-1&0&0\\
0&+2&0\\
0&0&+2
\end{pmatrix}\,.
\end{align}
%
%
%
We find that the color interaction does not mix these bases, just as in the $0^{+(+)}$ cases.
Moreover, the color interaction splits the three bases into two bands.
The color-sextet base is more stable than the two color-triplet bases by $128.7~\text{MeV}$.
%
The CM interaction will further split the two color-triplet bases, and give the three-band structure in Fig.~\ref{fig:cccb+bbbc}(a).
The $bb\bar{b}\bar{c}$ tetraquarks are very similar to the $cc\bar{c}\bar{b}$ tetraquarks.
Here we also find the ground state has $J^{P}=0^{+}$, which is dominated by the color-sextet $\ket{(bb)_{0}^{6}(\bar{b}\bar{c})_{0}^{\bar{6}}}$ component.
For the $1^{+}$ state, the lightest state $T(bb\bar{b}\bar{c},15719.1,1^{+})$ is mostly composed of $\ket{(bb)_{0}^{6}(\bar{b}\bar{c})_{1}^{\bar{6}}}$ ($93.7\%$) bases, while the two higher states $T(bb\bar{b}\bar{c},15851.3,1^{+})$ and $T(bb\bar{b}\bar{c},15854.4,1^{+})$ are dominated by $\ket{(bb)_{1}^{\bar{3}}(\bar{b}\bar{c})_{1}^{3}}$ ($99.8\%$) and $\ket{(bb)_{1}^{\bar{3}}(\bar{b}\bar{c})_{0}^{3}}$ ($93.5\%$) components respectively.

To study their decay properties, we transform the $cc\bar{c}\bar{b}$ ($bb\bar{b}\bar{c}$) tetraquarks into the $c\bar{c}{\otimes}c\bar{b}$ ($b\bar{b}{\otimes}b\bar{c}$) configuration, and calculate their partial decay width ratios.
The corresponding results can be found in Tables~\ref{table:wavefunc:cccb:13x24}--\ref{table:R:bb.bc}.
From Table~\ref{table:R:bb.bc}, we find that the $T(bb\bar{b}\bar{c},15719.1,1^{+})$ state does not have $S$-wave decay channel.
Moreover, the $\eta_{b}\bar{B}_{c}$ channel is also forbidden because of the conservation of the angular momentum and parity.
%
Thus we conclude that the $T(bb\bar{b}\bar{c},15719.1,1^{+})$ tetraquark is a narrow state.

One might wonder whether the conclusion will still hold if considering the following factors.
Owing to the lack of experimental data, we use the predicted $B_{c}^{*}$ mass of Ref.~\cite{Godfrey:1985xj} to estimate the dimeson thresholds.
Moreover, the $T(bb\bar{b}\bar{c},15719.1,1^{+})$ state is only {16.1~MeV} (17.9~MeV) below the $\Upsilon\bar{B}_{c}$ ($\eta_{b}\bar{B}_{c}^{*}$) threshold.
The uncertainty of the present model may be large enough to push the state upward, then this state may have an $S$-wave decay channel.
%
However, we note that the predicted $B_{c}^{*}$ mass was also used in parameter extraction, thus the error may be partially cancelled with each other.
Moreover, even if the mass of this state is pushed upward above the $\Upsilon\bar{B}_{c}$/$\eta_{b}\bar{B}_{c}^{*}$ threshold, its phase space is still relatively small.
Then the decay width of this state is still relatively small compared to that of other states.
%
%
We hope that future experiment can search for this state.
%

\subsection{The $cc\bar{b}\bar{b}$ system}
\label{sec:ccbb}

\begin{figure*}
\includegraphics[width=400pt]{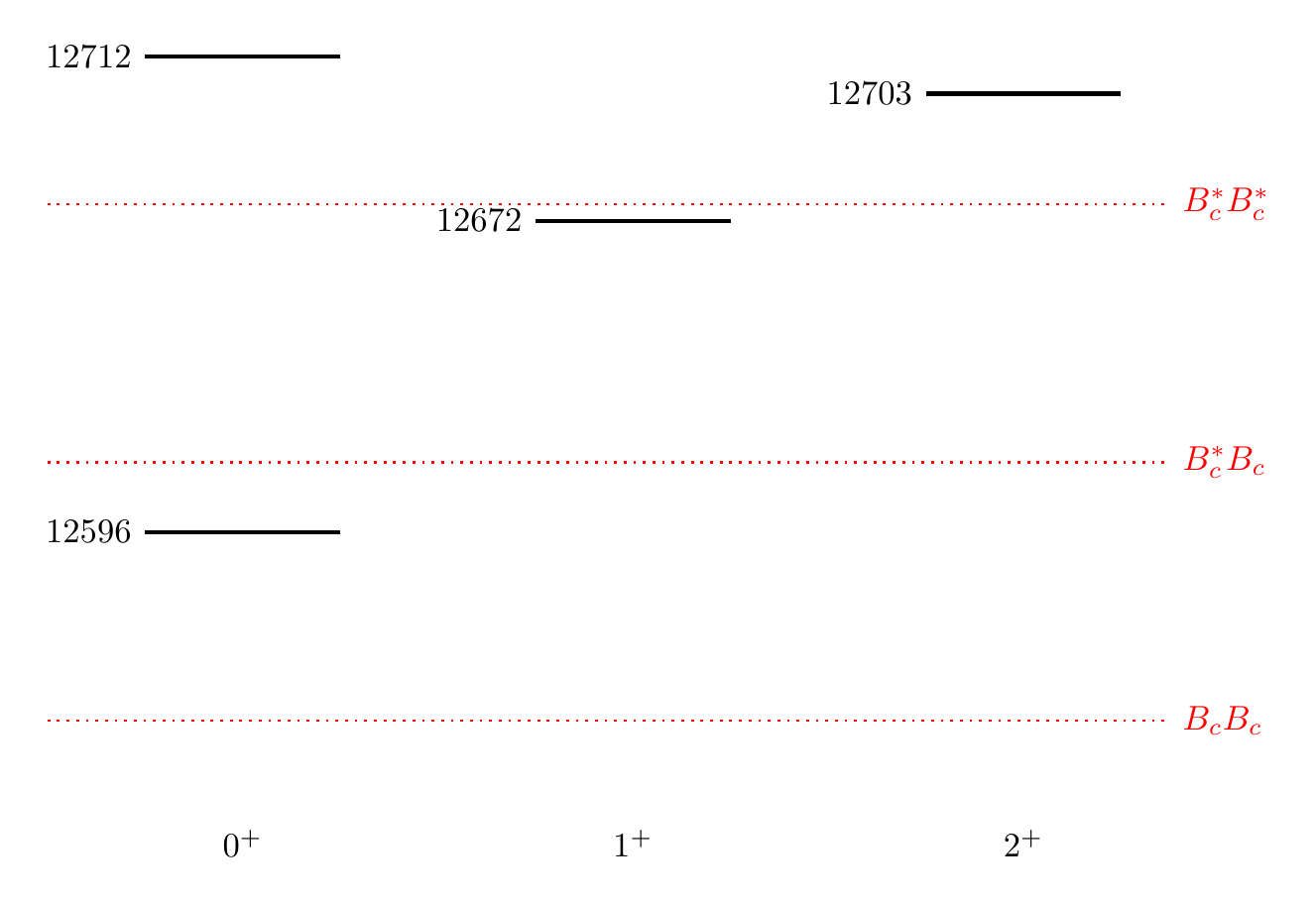}
\caption{Mass spectra of the $cc\bar{b}\bar{b}$ tetraquark states. The dotted lines indicate various meson-meson thresholds. Here the predicted mass $M_{B_{c}^{*}}=6338~\text{MeV}$ of Godfrey {\it et~al.}~\cite{Godfrey:1985xj} is used. The masses are all in units of MeV.}
\label{fig:ccbb}
\end{figure*}
%
\begin{table}
	\centering
	\caption{Masses and eigenvectors of the $cc\bar{b}\bar{b}$ tetraquarks. The masses are all in units of MeV.}
	\label{table:mass:ccbb}
	\begin{tabular}{ccccccc}
		\toprule[1pt]
		\toprule[1pt]
		System & $J^{P}$ & Mass & Eigenvector \\
		\midrule[1pt]
		$cc\bar{b}\bar{b}$&$0^{+}$&12596.3&$\{0.718,0.696\}$ \\
		&&12711.9&$\{-0.696,0.718\}$ \\
		&$1^{+}$&12671.7&$\{1\}$ \\
		&$2^{+}$&12703.1&$\{1\}$ \\
		\bottomrule[1pt]
		\bottomrule[1pt]
	\end{tabular}
\end{table}
%
\begin{table}
	\centering
	\caption{The eigenvectors of the $cc\bar{b}\bar{b}$ tetraquarks in the $c\bar{b}{\otimes}c\bar{b}$ configuration. The masses are all in units of MeV.}
	\label{table:wavefunc:ccbb:13x24}
	\begin{tabular}{ccccccc}
		\toprule[1pt]
		\toprule[1pt]
		System & $J^{P}$ & Mass & $B_{c}^{*}B_{c}^{*}$ & $B_{c}^{*}B_{c}$ & $B_{c}B_{c}^{*}$ & $B_{c}B_{c}$ \\
		\midrule[1pt]
		$cc\bar{b}\bar{b}$&$0^{+}$&12596.3&0.307&&&0.641 \\
		&&12711.9&$-0.699$&&&0.075 \\
		&$1^{+}$&12671.7&&0.408&0.408 \\
		&$2^{+}$&12703.1&0.577 \\
		\bottomrule[1pt]
		\bottomrule[1pt]
	\end{tabular}
\end{table}
%
\begin{table}
	\centering
	\caption{The values of $k\cdot|c_{i}|^2$ for the $cc\bar{b}\bar{b}$ tetraquarks (in units of MeV).}
	\label{table:kc_i^2:ccbb}
	\begin{tabular}{ccccccccccccc}
		\toprule[1pt]
		\toprule[1pt]
		System & $J^{PC}$ & Mass & $B_{c}^{*}B_{c}^{*}$ & $B_{c}^{*}B_{c}+B_{c}B_{c}^{*}$ & $B_{c}B_{c}$ \\
		\midrule[1pt]
		$cc\bar{b}\bar{b}$&$0^{+}$&12596.3&$\times$&&222.2 \\
		&&12711.9&233.5&&5.7 \\
		&$1^{+}$&12671.7&&203.1 \\
		&$2^{+}$&12703.1&138.2 \\
		\bottomrule[1pt]
		\bottomrule[1pt]
	\end{tabular}
\end{table}
%
\begin{table}
	\centering
	\caption{The partial width ratios for the $cc\bar{b}\bar{b}$ tetraquarks. For each state, we chose one mode as the reference channel, and the partial width ratios of the other channels are calculated relative to this channel. The masses are all in units of MeV.}
	\label{table:R:ccbb}
	\begin{tabular}{ccccccccccccc}
		\toprule[1pt]
		\toprule[1pt]
		System & $J^{PC}$ & Mass & $B_{c}^{*}B_{c}^{*}$ & $B_{c}^{*}B_{c}+B_{c}B_{c}^{*}$ & $B_{c}B_{c}$ \\
		\midrule[1pt]
		$cc\bar{b}\bar{b}$&$0^{+}$&12596.3&$\times$&&1 \\
		&&12711.9&41.0&&1 \\
		&$1^{+}$&12671.7&&1 \\
		&$2^{+}$&12703.1&1 \\
		\bottomrule[1pt]
		\bottomrule[1pt]
	\end{tabular}
\end{table}

We list the numerical results of the $cc\bar{b}\bar{b}$ tetraquark in Table~\ref{table:mass:ccbb}.
We also plot the mass spectrum and relevant meson-meson thresholds in Fig.~\ref{fig:ccbb}.
The pattern of mass spectrum is very similar to that of the $cc\bar{c}\bar{c}$/$bb\bar{b}\bar{b}$ tetraquarks.
The ground state and most heavy state both have quantum number $0^{+}$.
Their masses are $12596.3~\text{MeV}$ and $12711.9~\text{MeV}$ respectively.
The corresponding splitting is $115.6~\text{MeV}$, which is smaller than that of both $cc\bar{c}\bar{c}$ ($226.4~\text{MeV}$) and $bb\bar{b}\bar{b}$ ($144.9~\text{MeV}$) tetraquarks.
The reason is as follows.
With different quark/antiquark flavors, the color interaction becomes
\begin{equation}\label{eqn:HCE:ccnn}
\Braket{H_{\text{C}}\left(cc\bar{b}\bar{b}\right)}
=
2m_{c\bar{b}}
+
{\delta}\tilde{m}
\begin{pmatrix}
-1&0\\
0&+2
\end{pmatrix}
\end{equation}
where
\begin{equation}
{\delta}\tilde{m}
=
\frac{1}{2}
\left(
\frac{m_{cc}+m_{bb}}{2}
-
m_{c\bar{b}}
\right)
=
14.15~\text{MeV}\,,
\end{equation}
which is much smaller than the ${\delta}m_{Q}$ of Sec.~\ref{sec:cccc+bbbb}.
Thus the splitting becomes smaller.
%
%
%
Another consequence is that the CM interaction becomes relatively more important, thus the $\bar{3}_{c}{\otimes}3_{c}$ component becomes more important in the ground state.
This can be seen from the wave function listed in Table~\ref{table:mass:ccbb}, where the $\bar{3}_{c}{\otimes}3_{c}$ component of the ground state increases to $48.4\%$.
If we increase the mass difference between the quark and antiquark (such as $qq\bar{Q}\bar{Q}$ tetraquarks), the dominance of the $6_{c}{\otimes}\bar{6}_{c}$ or $\bar{3}_{c}{\otimes}{3}_{c}$ component may be reversed~\cite{Weng:2020zz01}.
%
%
%
Note that in some quark model studies, such reversing has already happened in the $cc\bar{b}\bar{b}$ system.
In Ref.~\cite{Wang:2019rdo}, Wang {\it et~al.} studied the $QQ\bar{Q}'\bar{Q}'$ tetraquark with two different quark model.
They found that in the model II, the ground state has $53\%$ of the $\bar{3}_{c}{\otimes}3_{c}$ component, which is slightly larger than the $6_{c}{\otimes}\bar{6}_{c}$ component.
%
%
%
%
A detailed study of the dependence of the mass spectrum and wave function with respect to the mass ratio $m_{q}/m_{\bar{q}}$ is very important to reveal the nature of tetraquarks.
%

%

Next we consider their decay properties.
We transform the wave functions of the $cc\bar{b}\bar{b}$ tetraquark into the $c\bar{b}{\otimes}c\bar{b}$ configuration, as shown in Table~\ref{table:wavefunc:ccbb:13x24}.
Experimentally, only the $B_{c}$ states have been found~\cite{Zyla:2020zbs}.
%
Here we use the $M_{B_{c}^{*}}=6338~\text{MeV}$ predicted by the Godfrey-Isgur model~\cite{Godfrey:1985xj}, as we did in Ref.~\cite{Weng:2018mmf} to estimate the model parameters.
The $k\cdot|c_{i}|^2$ values and relative decay widths are listed in Tables~\ref{table:kc_i^2:ccbb}--\ref{table:R:ccbb}, where we have assumed
\begin{equation}
\gamma_{B_{c}^{*}B_{c}^{*}}
=
\gamma_{B_{c}^{*}B_{c}}
=
\gamma_{B_{c}B_{c}^{*}}
=
\gamma_{B_{c}B_{c}}\,.
\end{equation}
Similar to the $cc\bar{c}\bar{c}$/$bb\bar{b}\bar{b}$ cases, we find that all states are above the thresholds into two $S$-wave mesons.
Thus they are broad~\cite{Jaffe:1976ig}.
%
The ground state can only decay into two $B_{c}$ mesons.
And the other scalar state $T(cc\bar{b}\bar{b},12711.9,0^{+})$ decays into both ${B_{c}B_{c}}$ and ${B_{c}^{*}B_{c}^{*}}$ modes.
Their partial decay width ratio is
\begin{equation}
\frac{\Gamma\left[T(cc\bar{b}\bar{b},12711.9,0^{++}){\to}{B_{c}^{*}B_{c}^{*}}\right]}{\Gamma\left[T(cc\bar{b}\bar{b},12711.9,0^{++}){\to}{B_{c}B_{c}}\right]}
=
41.0\,.
\end{equation}
Thus the ${B_{c}^{*}B_{c}^{*}}$ mode is dominant.
The other two states $T(cc\bar{b}\bar{b},12671.7,1^{+})$ and $T(cc\bar{b}\bar{b},12703.1,2^{+})$ can decay into ${B_{c}B_{c}^{*}}$ and ${B_{c}^{*}B_{c}^{*}}$ modes respectively in $S$-wave.
%
%

\subsection{The $cb\bar{c}\bar{b}$ system}
\label{sec:cbcb}

\begin{figure*}
\includegraphics[width=400pt]{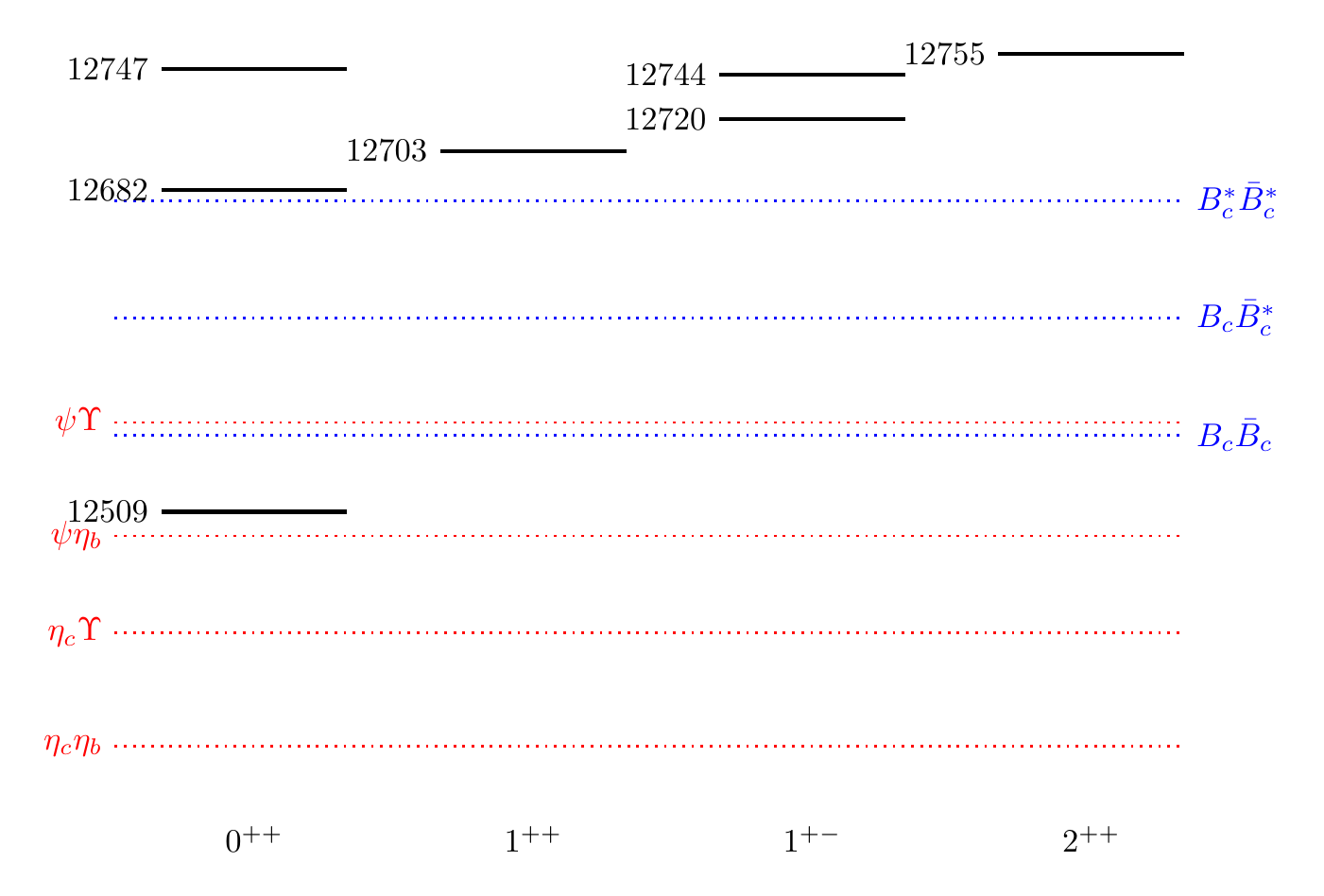}
\caption{Mass spectra of the $cb\bar{c}\bar{b}$ tetraquark states. The dotted lines indicate various meson-meson thresholds. Here the predicted mass $M_{B_{c}^{*}}=6338~\text{MeV}$ of Godfrey {\it et~al.}~\cite{Godfrey:1985xj} is used. The masses are all in units of MeV.}
\label{fig:cbcb}
\end{figure*}
%
\begin{table*}
	\centering
	\caption{Masses and eigenvectors of the $cb\bar{c}\bar{b}$ tetraquarks. The masses are all in units of MeV.}
	\label{table:mass:cbcb}
	\begin{tabular}{ccccccc}
		\toprule[1pt]
		\toprule[1pt]
		System&$J^{PC}$&Mass&Eigenvector&Scattering~state\\
		\midrule[1pt]
		$cb\bar{c}\bar{b}$&$0^{++}$&12362.8&$\{0.868,0.298,0.296,0.264\}$&$\eta_{c}\eta_{b}$\\
		&&12509.3&$\{-0.415,0.873,0.160,0.199\}$\\
		&&12681.6&$\{0.013,-0.043,-0.662,0.748\}$\\
		&&12746.9&$\{-0.272,-0.383,0.670,0.575\}$\\
		&$1^{++}$&12523.6&$\{0.697,0.697,0.118,0.118\}$&$\psi\Upsilon$\\
		&&12703.2&$\{-0.118,-0.118,0.697,0.697\}$\\
		&$1^{+-}$&12424.9&$\{0.794,-0.349,0.349,0.252,-0.176,0.176\}$&$\eta_{c}\Upsilon$\\
		&&12477.2&$\{0.539,0.535,-0.535,0.128,0.245,-0.245\}$&$\psi\eta_{b}$\\
		&&12720.0&$\{0.279,-0.002,0.002,-0.957,-0.053,0.053\}$\\
		&&12744.1&$\{0.035,-0.303,0.303,-0.058,0.637,-0.637\}$\\
		&$2^{++}$&12537.4&$\{0.953,0.304\}$&$\psi\Upsilon$\\
		&&12754.9&$\{-0.304,0.953\}$\\
		\bottomrule[1pt]
		\bottomrule[1pt]
	\end{tabular}
\end{table*}
%
\begin{table}
	\centering
	\caption{The eigenvectors of the $cb\bar{c}\bar{b}$ tetraquarks in the $c_{1}\bar{c}_{3}{\otimes}b_{2}\bar{b}_{4}$ configuration. The masses are all in units of MeV.}
	\label{table:wavefunc:cbcb:13x24}
	\begin{tabular}{ccccccc}
		\toprule[1pt]
		\toprule[1pt]
		System&$J^{PC}$&Mass&$\psi\Upsilon$&$\psi\eta_{b}$&$\eta_{c}\Upsilon$&$\eta_{c}\eta_{b}$\\
		\midrule[1pt]
		$cb\bar{c}\bar{b}$&$0^{++}$&12362.8&$-0.097$&&&$0.960$\\
		&&12509.3&$0.840$&&&$0.200$\\
		&&12681.6&$0.529$&&&$-0.123$\\
		&&12746.9&$-0.066$&&&$0.152$\\
		&$1^{++}$&12523.6&$-0.902$\\
		&&12703.2&$0.433$\\
		&$1^{+-}$&12424.9&&$0.174$&$0.948$\\
		&&12477.2&&$0.942$&$-0.215$\\
		&&12720.0&&$-0.261$&$-0.198$\\
		&&12744.1&&$0.117$&$-0.124$\\
		&$2^{++}$&12537.4&$0.953$\\
		&&12754.9&$0.302$\\
		\bottomrule[1pt]
		\bottomrule[1pt]
	\end{tabular}
\end{table}
%
\begin{table}
	\centering
	\caption{The eigenvectors of the $cb\bar{c}\bar{b}$ tetraquarks in the $c_{1}\bar{b}_{4}{\otimes}b_{2}\bar{c}_{3}$ configuration. The masses are all in units of MeV.}
	\label{table:wavefunc:cbcb:14x23}
	\begin{tabular}{ccccccc}
		\toprule[1pt]
		\toprule[1pt]
		System&$J^{PC}$&Mass&${B}_{c}^{*}\bar{B}_{c}^{*}$&${B}_{c}^{*}\bar{B}_{c}$&${B}_{c}\bar{B}_{c}^{*}$&${B}_{c}\bar{B}_{c}$\\
		\midrule[1pt]
		$cb\bar{c}\bar{b}$&$0^{++}$&12362.8&$-0.348$&&&$0.420$\\
		&&12509.3&$-0.302$&&&$-0.673$\\
		&&12681.6&$0.208$&&&$0.574$\\
		&&12746.9&$0.863$&&&$-0.205$\\
		&$1^{++}$&12523.6&&$0.501$&$-0.501$\\
		&&12703.2&&$0.499$&$-0.499$\\
		&$1^{+-}$&12424.9&$0.259$&$0.355$&$0.355$\\
		&&12477.2&$-0.418$&$0.259$&$0.259$\\
		&&12720.0&$-0.041$&$0.552$&$0.552$\\
		&&12744.1&$0.870$&$0.044$&$0.044$\\
		&$2^{++}$&12537.4&$0.602$\\
		&&12754.9&$-0.798$\\
		\bottomrule[1pt]
		\bottomrule[1pt]
	\end{tabular}
\end{table}
%
\begin{table}
	\centering
	\caption{The partial width ratios for the $cb\bar{c}\bar{b}$ tetraquarks decay into a charmonium plus a bottomonium. For each state, we chose one mode as the reference channel, and the partial width ratios of the other channels are calculated relative to this channel. The masses are all in units of MeV.}
	\label{table:R:cbcb:13x24}
	\begin{tabular}{ccccccccccccc}
		\toprule[1pt]
		\toprule[1pt]
		System&$J^{PC}$&Mass&$\psi\Upsilon$&$\psi\eta_{b}$&$\eta_{c}\Upsilon$&$\eta_{c}\eta_{b}$\\
		\midrule[1pt]
		$cb\bar{c}\bar{b}$&$0^{++}$&12509.3&$\times$&&&1\\
		&&12681.6&12.0&&&1\\
		&&12746.9&0.1&&&1\\
		&$1^{++}$&12703.2&1\\
		&$1^{+-}$&12720.0&&1.6&1\\
		&&12744.1&&0.8&1\\
		&$2^{++}$&12754.9&1\\
		\bottomrule[1pt]
		\bottomrule[1pt]
	\end{tabular}
\end{table}
%
\begin{table}
	\centering
	\caption{The partial width ratios for the $cb\bar{c}\bar{b}$ tetraquarks decays into $B_{c}^{(*)}\bar{B}_{c}^{(*)}$. For each state, we chose one mode as the reference channel, and the partial width ratios of the other channels are calculated relative to this channel. The masses are all in units of MeV.}
	\label{table:R:cbcb:14x23}
	\begin{tabular}{ccccccccccccc}
		\toprule[1pt]
		\toprule[1pt]
		System&$J^{PC}$&Mass&${B}_{c}^{*}\bar{B}_{c}^{*}$&${B}_{c}^{*}\bar{B}_{c}$&${B}_{c}\bar{B}_{c}^{*}$&${B}_{c}\bar{B}_{c}$\\
		\midrule[1pt]
		$cb\bar{c}\bar{b}$&$0^{++}$&12509.3&$\times$&&&$\times$\\
		&&12681.6&0.03&&&1\\
		&&12746.9&10.7&&&1\\
		&$1^{++}$&12703.2&&1&1\\
		&$1^{+-}$&12720.0&0.003&1&1\\
		&&12744.1&278.7&1&1\\
		&$2^{++}$&12754.9&1\\
		\bottomrule[1pt]
		\bottomrule[1pt]
	\end{tabular}
\end{table}

Now we turn to the $cb\bar{c}\bar{b}$ system.
The mass spectra and eigenvectors are listed in Table~\ref{table:mass:cbcb}.
We also transform the eigenvectors into the $c\bar{c}{\otimes}b\bar{b}$ and $c\bar{b}{\otimes}b\bar{c}$ configurations, as shown in Tables.~\ref{table:wavefunc:cbcb:13x24}--\ref{table:wavefunc:cbcb:14x23}.
%
%
From Table~\ref{table:wavefunc:cbcb:13x24}, we see that the eigenvector of the lowest state reads
\begin{equation}
T(cb\bar{c}\bar{b},12362.8,0^{++})
=
0.960\eta_{c}\eta_{c}
+\cdots\,.
\end{equation}
This state couples very strongly to the $\eta_{c}\eta_{c}$ channel.
It is broad and is just a part of the continuum.
Note that this kind of state also exists in our previous study of the hidden charm pentaquark, where some of the calculated states couple strongly to a charmonium and a light baryons~\cite{Weng:2019ynv}.
%
%
Similar phenomenon has also been found in $c\bar{c}q\bar{q}$ tetraquark~\cite{Cui:2006mp,Hogaasen:2013nca}.
Moreover, the states of $12523.6~\text{MeV}$ (with $J^{PC}=1^{++}$) and $12537.4~\text{MeV}$ (with $J^{PC}=2^{++}$) couple strongly to $\psi$ and $\Upsilon$ mesons.
They are also scattering states.
The remaining coupling type of $\psi\otimes\Upsilon$ is $0^{++}$, which largely resides in $T(cb\bar{c}\bar{b},12509.3,0^{++})$.
However, this state possesses large fractions ($25\%$) of the color-octet components, thus we cannot rule out the possibility that it is a tetraquark.
To draw a more definitive conclusion, we need to study its internal dynamics~\cite{Wang:2019rdo,Myo:2020rni,Yang:2020atz}, which is beyond the present work.
%
%
There are two additional scattering states composed of a vector meson and a pseudoscalar meson.
The $T(cb\bar{c}\bar{b},12424.9,1^{+-})$ is a scattering state of $\eta_{c}\Upsilon$ and $T(cb\bar{c}\bar{b},12477.2,1^{+-})$ is a scattering state of $\psi\eta_{b}$.
For clarity, we add a fifth column in Table~\ref{table:mass:cbcb} to indicate these scattering states.

In Fig.~\ref{fig:cbcb}, we present the relative position of the $cb\bar{c}\bar{b}$ tetraquarks.
We also plot all the relevant meson-meson thresholds.
%
%
%
To study their decay properties, we need to estimate the $\gamma_{i}$.
In the quark model, the spatial wave functions of the ground state scalar and vector meson are the same.
Thus for each $cb\bar{c}\bar{b}$ tetraquark
\begin{equation}
\gamma_{{\psi}\Upsilon}
=
\gamma_{{\psi}{\eta_{b}}}
=
\gamma_{{\eta_{c}}\Upsilon}
=
\gamma_{{\eta_{c}}\eta_{b}}
\end{equation}
and
\begin{equation}
\gamma_{B_{c}^{*}B_{c}^{*}}
=
\gamma_{B_{c}^{*}B_{c}}
=
\gamma_{B_{c}B_{c}^{*}}
=
\gamma_{B_{c}B_{c}}\,.
\end{equation}
Combining the eigenvectors in the $c\bar{c}{\otimes}b\bar{b}$ and $c\bar{b}{\otimes}b\bar{c}$ configurations, we can calculate the relative  partial widths of different decay modes, as listed in Tables~\ref{table:R:cbcb:13x24}--\ref{table:R:cbcb:14x23}.

The lowest $cb\bar{c}\bar{b}$ tetraquarks is $T(cb\bar{c}\bar{b},12509.3,0^{++})$.
This state can decay into $\eta_{c}\eta_{b}$ through $S$-wave.
The $T(cb\bar{c}\bar{b},12681.6,0^{++})$ and $T(cb\bar{c}\bar{b},12746.9,0^{++})$ tetraquarks have the same decay channels.
But their relative sizes of partial width are different, which can be used to distinguish the two states.
More precisely, for $T(cb\bar{c}\bar{b},12681.6,0^{++})$, we have
\begin{equation}
	\Gamma_{\psi\Upsilon}:\Gamma_{\eta_{c}\eta_{b}}
	=
	12.0\,,
\end{equation}
and
\begin{equation}
	\Gamma_{B_{c}^{*}\bar{B}_{c}^{*}}:\Gamma_{B_{c}\bar{B}_{c}}
	=
	0.03\,.
\end{equation}
And the $T(cb\bar{c}\bar{b},12746.9,0^{++})$ has
\begin{equation}
	\Gamma_{\psi\Upsilon}:\Gamma_{\eta_{c}\eta_{b}}
	=
	0.1\,,
\end{equation}
and
\begin{equation}
	\Gamma_{B_{c}^{*}\bar{B}_{c}^{*}}:\Gamma_{B_{c}\bar{B}_{c}}
	=
	10.7\,.
\end{equation}
The other states (with $J=1$ and $J=2$) are all above the $B_{c}^{*}\bar{B}_{c}^{*}$ threshold.
They may be broad since they can freely decay into many channels in $S$-wave.
%

\section{Conclusions}
\label{Sec:Conclusion}

In this work, we have systematically studied the mass spectrum of the fully heavy tetraquark in an extended chromomagnetic model, which includes both colorelectric and chromomagnetic interactions.
There is no stable state below the lowest heavy quarkonium pair thresholds.
Most states can dissociate into two $S$-wave mesons through $S$-wave decay.
Thus they may be wide states.
One possible narrow state is the $1^{+}$ $bb\bar{b}\bar{c}$ state with mass $15719.1~\text{MeV}$.
Although it is above the $\eta_{b}\bar{B}_{c}$ threshold, this channel is forbidden because of the conservation of the angular momentum and parity.

There are two possible color configurations of tetraquark, namely the color-sextet $\ket{(QQ)^{6_{c}}(\bar{Q}\bar{Q})^{\bar{6}_{c}}}$ and the color-triplet $\ket{(QQ)^{\bar{3}_{c}}(\bar{Q}\bar{Q})^{3_{c}}}$.
From the eigenvectors obtained, we find that the energy level is mainly determined by the colorelectric interaction.
The colorelectric interaction always favors the color-sextet configurations.
Note that if $Q_{1}=Q_{2}$ or $\bar{Q}_{3}=\bar{Q}_{4}$, the colorelectric interaction does not mix the two color configurations.
%
%
The chromomagnetic interaction favors the color-triplet configurations.
But its contribution is relatively smaller than that of the colorelectric interaction, thus it only gives small splitting to the states.
The chromomagnetic interaction can also mix the color-sextet and the color-triplet configurations.
Comparing the ground states [$J^{PC}=0^{+(+)}$] of the $cc\bar{c}\bar{c}$/$bb\bar{b}\bar{b}$ and $cc\bar{b}\bar{b}$ tetraquarks, we find that the mixing becomes larger when there is a mass difference between the quark and antiquark.
In other words, the color-triplet component is more important in the ground state $cc\bar{b}\bar{b}$ tetraquark than in the ground state $cc\bar{c}\bar{c}$/$bb\bar{b}\bar{b}$ tetraquarks.
%
%

With the eigenvectors obtained, we have also investigated the decay properties of the tetraquarks.
We hope that future experiments can search for these states.
%

\section*{Acknowledgments}

XZW is grateful to G.~J.~Wang and L.~Meng for helpful comments and discussions.
This project is supported by the National Natural Science Foundation of China under Grants No. 11975033.
%



%
%

\bibliography{myreference}
\end{document}